\documentclass[aps,pra,twocolumn,amsfonts,amssymb,amsmath,showpacs,floatfix,nofootinbib,groupedaddress,superscriptaddress,citesort]{revtex4}
\usepackage{graphics,graphicx,times,bm,multirow,color,amstext,theorem}

\newtheorem{theorem}{Theorem}
{\theorembodyfont{\rmfamily}
\newtheorem{lemma}{Lemma}
\newtheorem{definition}{Definition}
\newtheorem{condition}{Condition}
 \theoremstyle{break}}
\newenvironment{proof}{{\bf Proof: \ }}{ \hfill}

\begin{document}

\title{Direct Characterization of Quantum Dynamics: General Theory}
\author{M. Mohseni}
\affiliation{Department of Chemistry and Chemical Biology, Harvard
University, 12 Oxford St., Cambridge, MA 012138}
\affiliation{Department of Chemistry, University of Southern
  California, Los Angeles, CA 90089}
\author{D. A. Lidar}
\affiliation{Department of Chemistry, University of Southern
  California, Los Angeles, CA 90089}
\affiliation{Departments of Electrical Engineering and Physics, University of Southern
California, Los Angeles, CA 90089}

\begin{abstract}

The characterization of the dynamics of quantum systems is a
task of both fundamental and practical importance. A general class of
methods which have been developed in quantum
information theory to accomplish this task is known as quantum process tomography (QPT). In an earlier paper
[M. Mohseni and D. A. Lidar, Phys. Rev. Lett. \textbf{97}, 170501
  (2006)] we presented a new algorithm for Direct Characterization of
Quantum Dynamics (DCQD) of two-level quantum systems. Here we provide a generalization by developing a theory for direct and
complete characterization of the dynamics of arbitrary
quantum systems. In contrast to other QPT schemes, DCQD relies on
quantum error-detection techniques and does not require
any quantum state tomography. We demonstrate that for the full
characterization of the dynamics of $n$ $d$-level quantum systems (with
$d$ a power of a prime), the minimal number of required
experimental configurations is reduced quadratically from $d^{4n}$
in separable QPT schemes to $d^{2n}$ in DCQD.
\end{abstract}

\maketitle


\section{Introduction}

\label{intro}

The characterization of quantum dynamical systems is a
fundamental problem in quantum physics and quantum chemistry. Its
ubiquity is due to the fact that knowledge of quantum dynamics of (open or closed) quantum systems is
indispensable in prediction of experimental outcomes. In particular, accurate
estimation of an unknown quantum dynamical process acting on a
quantum system is a pivotal task in coherent control of the
dynamics, especially in verifying/monitoring the performance of a
quantum device in the presence of decoherence. The procedures for
characterization of quantum dynamical maps are traditionally known
as quantum process tomography (QPT)
\cite{nielsen-book,d'ariano-qt,mohseni-rezakhani-lidar07}.

In most QPT schemes the information about the quantum dynamical
process is obtained indirectly. The quantum dynamics is first mapped
onto the state(s) of an ensemble of probe quantum systems, and then
the process is reconstructed via quantum state tomography of the
output states. Quantum state tomography is itself a procedure for
identifying a quantum system by measuring the expectation values of
a set of non-commuting observables on identical copies of the
system. There are two general types QPT schemes. The first is
Standard Quantum Process Tomography (SQPT)
\cite{nielsen-book,chuang-sqpt,poyatos-sqpt}. In SQPT all quantum
operations, including preparation and (state tomography)
measurements, are performed on the system whose dynamics is to be
identified (the ``principal'' system), without the use of any
ancillas. The SQPT scheme has already been experimentally
demonstrated in a variety of systems including liquid-state nuclear
magnetic resonance (NMR) \cite{childs-nmr,cory03,cory04}, optical
\cite{steinberg-bell,james-white}, atomic \cite{steinberg-qpt2}, and
solid-state systems \cite{howard}. The second type of QPT scheme is
known as Ancilla-Assisted Process Tomography (AAPT)
\cite{d'ariano-aapt,leung,altepeter-aapt,d'ariano-faithful}. In AAPT
one makes use of an ancilla (auxilliary system). First, the combined
principal system and ancilla are prepared in a ``faithful'' state,
with the property that all information about the dynamics can be
imprinted on the final state
\cite{d'ariano-aapt,altepeter-aapt,d'ariano-faithful}. The relevant
information is then extracted by performing quantum state tomography
in the joint Hilbert space of system and ancilla. The AAPT scheme
has also been demonstrated experimentally
\cite{Martini,altepeter-aapt}. The total number of experimental
configurations required for measuring the quantum dynamics of $n$
$d$-level quantum systems (qudits) is $d^{4n}$ for both SQPT and
separable AAPT, where separable refers to the measurements performed
at the end. This number can in principle be reduced by utilizing
non-separable measurements, e.g., a generalized measurement
\cite{nielsen-book}. However, the non-separable QPT schemes are
rather impractical in physical applications because they require
many-body interactions, which are not experimentally available or
must be simulated at high resource cost
\cite{mohseni-rezakhani-lidar07}.

Both SQPT and AAPT make use of a mapping of the dynamics onto a
state. This raises the natural question of whether it is possible to
avoid such a mapping and instead perform a \emph{direct} measurement
of quantum dynamics, which does not require any state tomography.
Moreover, it seems reasonable that by avoiding the indirect mapping
one should be able to attain a reduction in resource use (e.g., the
total number of measurements required), by eliminating redundancies.
Indeed, there has been a growing interest in the development of
direct methods for obtaining specific information about the states
or dynamics of quantum systems. Examples include the estimation of
general functions of a quantum state \cite{ekert-direct}, detection
of quantum entanglement \cite{horodecki-direct}, measurement of
nonlinear properties of bipartite quantum states
\cite{bovino-direct}, reconstruction of quantum states or dynamics
from incomplete measurements \cite{ziman}, estimation of the average
fidelity of a quantum gate or process
\cite{Emerson-direct,Hofmann-direct}, and universal source coding
and data compression \cite{bennett-compression}. However, these
schemes cannot be used directly for a \emph{complete}
characterization of quantum dynamics. In Ref.~\cite{mohseni-lidar06}
we presented such a scheme, which we called ``Direct
Characterization of Quantum Dynamics'' (DCQD).

\begin{figure}[tbp]
\includegraphics[width=7cm,height=1.8cm]{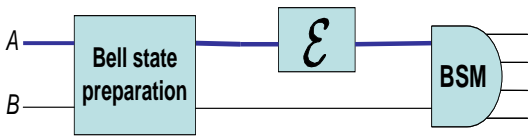}
\caption{Schematic of DCQD for a single qubit, consisting of
  Bell-state preparations, application of the unknown quantum map,
$\mathcal{E}$, and Bell-state measurement (BSM).} \label{bsmf}
\end{figure}

In trying to address the problem of \emph{direct} and \emph{complete}
characterization of quantum dynamics, we were
inspired by the observation that quantum error detection (QED)
\cite{nielsen-book} provides a means to directly obtain
partial information about the nature of a quantum process, without
ever revealing the state of the system. In
general, however, it is unclear if there is a fundamental
relationship between QED and QPT, namely whether it is possible to completely
characterize the quantum dynamics of arbitrary quantum systems
using QED. And, providing the answer is affirmative, how the
physical resources scale with system size.
Moreover, one would like to understand whether entanglement plays a
fundamental role, and what potential applications emerge from
such a theory linking QPT and QED. Finally, one would hope that this approach may lead to new ways of
understanding and/or controlling quantum dynamical systems. We
addressed these questions for the first time in
Ref.~\cite{mohseni-lidar06} by developing the DCQD algorithm in the
context of two-level quantum systems. In DCQD -- see Fig.~\ref{bsmf} --
the state space of an ancilla is utilized such that
experimental outcomes from a Bell-state measurement provide direct
information about specific properties of the underlying dynamics. A
complete set of probe states is then used to fully characterize the
unknown quantum dynamics via application of a single Bell-state
measurement device \cite{mohseni-lidar06,mohseni-rezakhani-lidar07}.

Here we generalize the theory of
Ref.~\cite{mohseni-lidar06} to arbitrary open quantum systems
undergoing an unknown, completely-positive (CP) quantum dynamical
map. In the generalized DCQD scheme, each probe qudit (with $d$ prime) is
initially entangled with an ancillary qudit system of the same
dimension, before being subjected to the unknown quantum process. To
extract the relevant information, the corresponding measurements are
devised in such a way that the final (joint) probability
distributions of the outcomes are directly related to specific sets of the dynamical superoperator's elements. A complete set of
probe states can then be utilized to fully characterize the unknown
quantum dynamical map. The preparation of the probe systems and the
measurement schemes are based on QED techniques, however, the
objective and the details of the error-detection schemes are
different from those appearing in the protection of quantum systems against
decoherence (the original context of QED). More specifically, we develop error-detection schemes to
directly measure the coherence in a quantum dynamical process,
represented by off-diagonal elements of the corresponding
superoperator. We explicitly demonstrate that for characterizing a
dynamical map on $n$ qudits, the number of required experimental
configurations is reduced from $d^{4}{}^{n}$, in SQPT and
separable AAPT, to $d^{2n}$ in DCQD. A useful feature of DCQD is that it can be
efficiently applied to \textit{partial} characterization of quantum
dynamics \cite{mohseni-lidar06,MasoudThesis}. For example, it can be used for the task of Hamiltonian
identification, and also for simultaneous determination of the
relaxation time $T_{1}$ and the dephasing time $T_{2}$.

This paper is organized as follows. In Sec.~\ref{preliminaries}, we
provide a brief review of completely-positive quantum dynamical
maps, and the relevant QED concepts such as stabilizer codes and
normalizers. In Sec.~\ref{Population-qudit}, we demonstrate how to
determine the quantum dynamical populations, or diagonal elements of
a superoperator, through a single (ensemble) measurement. In order to
further develop the DCQD algorithm and build the required notations,
we introduce some lemmas and definitions in Sec.~\ref{basic-lemmas},
and then we address the characterization of quantum dynamical
coherences, or
off-diagonal elements of a superoperator, in Sec.~\ref{coherence-qudit}. In Sec.~\ref%
{linearindep-optimality}, we show that measurement outcomes obtained in Sec.~\ref%
{coherence-qudit} provide $d^{2}$ linearly independent equations for
estimating the coherences in a process, which is in fact the maximum
amount of information that can be extracted in a single measurement.
A complete characterization of the quantum dynamics, however,
requires obtaining $d^{4}$ independent real parameters of the superoperator (for non-trace preserving maps). In Sec.~\ref%
{rep-algorithm}, we demonstrate how one can obtain complete
information by appropriately rotating the input state and repeating
the above algorithm for a complete set of rotations. In Sec.~\ref{constraints-prep} and \ref%
{minimum-qudits}, we address the general constraints on input
stabilizer codes and the minimum number of physical qudits required
for the encoding. In Sec.~\ref{standar-form-S} and
Sec.~\ref{algorithm-summary}, we define a standard notation for
stabilizer and normalizer measurements and then provide an outline
of the DCQD algorithm for the case of a single qudit. For convenience,
we provide a brief summary of the entire DCQD algorithm in
Sec.~\ref{summary-qudit}. We conclude with an outlook in
Section~\ref{outlook}. In
Appendix~\ref{generalization-Algorithm}, we generalize the scheme
for arbitrary open quantum systems. For a discussion of the
experimental feasibility of DCQD see Ref.~\cite{mohseni-lidar06},
and for a detailed and comprehensive comparison of the required
physical resources in different QPT schemes see
Ref.~\cite{mohseni-rezakhani-lidar07}.

\section{Preliminaries}

\label{preliminaries}

In this section we introduce the basic concepts and notation from
the theory of open quantum system dynamics and quantum error
detection, required for the generalization of the DCQD algorithm to
qudits.

\subsection{Quantum Dynamics}

The evolution of a quantum system (open or closed) can, under
natural assumptions, be expressed in terms of a completely positive
quantum dynamical map $\mathcal{E}$, which can be represented as
\cite{nielsen-book}
\begin{equation}
{\mathcal{E}}(\rho )=\sum_{m,n=0}^{d^{2}-1}\chi _{mn}~E_{m}\rho
E_{n}^{\dagger }.
\end{equation}
Here $\rho $ is the initial state of the system, and the $\{E_{m}\}$ are a
set of (error) operator basis elements in the Hilbert-Schmidt space
of the linear operators acting on the system. I.e., any arbitrary
operator acting on a $d$-dimensional quantum system can
be expanded over an orthonormal and unitary error operator basis $%
\{E_{0},E_{1},\ldots,E_{d^{2}-1}\}$, where $E_{0}=I$ and \textrm{tr}$%
(E_{i}^{\dagger }E_{j})=d\delta _{ij}$ \cite{Knillhighd}. The
$\{\chi _{mn}\}$ are the matrix elements of the superoperator
$\bm{\chi}$, or ``process matrix'', which encodes all the information about the dynamics,
relative to the basis set $\{E_{m}\}$ \cite{nielsen-book}. For an
$n$-qudit system, the number of independent matrix elements in
$\bm{\chi}$ is $d^{4n}$ for a non-trace-preserving map and $d^{4n}-d^{2n}$ for a
trace-preserving map. The process matrix $\bm{\chi}$ is positive and $%
\mathrm{Tr}\bm{\chi}\leq 1$. Thus $\bm{\chi}$ can be thought of as a
density matrix in the Hilbert-Schmidt space, whence we often refer to its diagonal and off-diagonal elements as
\textquotedblleft quantum dynamical population\textquotedblright\
and \textquotedblleft quantum dynamical coherence\textquotedblright
, respectively.

In general, any successive operation of the (error) operator basis can be expressed as $%
E_{i}E_{j}=\sum_{k}\omega ^{i,j,k}E_{k}$, where
$i,j,k=0,1,\ldots,d^{2}-1$. However, we use the ``very nice (error) operator basis"
in which $E_{i}E_{j}=\omega ^{i,j}E_{i\ast j}$, $\det E_{i}=1$, $%
\omega ^{i,j}$ is a $d$th root of unity, and the operation $\ast $
induces a group on the indices \cite{Knillhighd}. This provides a
natural generalization of the Pauli group to higher dimensions. Any
element $E_{i}$
can be generated from appropriate products of $X_{d}$ and $Z_{d}$, where $%
X_{d}\left\vert k\right\rangle =\left\vert k+1\right\rangle $ , $%
Z_{d}\left\vert k\right\rangle =\omega ^{k}\left\vert k\right\rangle $, and $%
X_{d}Z_{d}=\omega ^{-1}Z_{d}X_{d}$
\cite{Gottessmanhighd,Knillhighd}. Therefore, for any two elements
$E_{i=\{a,q,p\}}=\omega ^{a}X_{d}^{q}Z_{d}^{p}$ and
$E_{j=\{a^{\prime },q^{\prime },p^{\prime
}\}}=\omega ^{a^{\prime }}X_{d}^{q^{\prime }}Z_{d}^{p^{\prime }}$ (where $%
0\leq q,p<d$) of the single-qudit Pauli group, we always have
\begin{equation}
E_{i}E_{j}=\omega ^{pq^{\prime }-qp^{\prime }}E_{j}E_{i}~,
\label{ErrorsCommutation}
\end{equation}%
where
\begin{equation}
pq^{\prime }-qp^{\prime }\equiv k~(\mathrm{mod}~d).
\label{EigenvalueCom}
\end{equation}
The operators $E_{i}$ and $E_{j}$ commute iff $k=0$. Henceforth, all algebraic operations are performed in $\mathrm{mod}%
(d)$ arithmetic, and all quantum states and operators,
respectively, belong to and act on a $d$-dimensional Hilbert space.
For simplicity, from now on we drop the subscript $d$ from the
operators.

\subsection{Quantum Error Detection}

In the last decade the theory of quantum error correction (QEC) has
been developed as a general method for detecting and correcting
quantum dynamical errors acting on multi-qubit systems such as a
quantum computer \cite{nielsen-book}. QEC consists of three steps:
preparation, quantum error detection (QED) or syndrome measurements,
and recovery. In the preparation step, the state of a quantum system
is encoded into a subspace of a larger Hilbert space by entangling
the principal system with some other quantum systems using unitary
operations. This encoding is designed to allow detection of
arbitrary errors on one (or more) physical qubits of a code by
performing a set of QED measurements. The measurement strategy is to
map different possible sets of errors only to orthogonal and
undeformed subspaces of the total Hilbert space, such that the
errors can be unambiguously discriminated. Finally the detected
errors can be corrected by applying the required unitary operations
on the physical qubits during the recovery step. A key observation
relevant for our purposes
is that by performing QED one can actually obtain partial
information about the dynamics of an open quantum system.

\begin{figure*}[tp]
\begin{center}
\includegraphics[width=12cm,height=4.1cm]{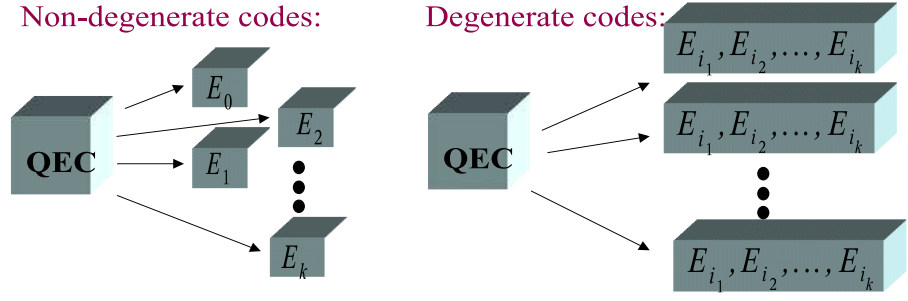}
\end{center}
\caption[A schematic diagram of Quantum Error Detection
(QED)]{\small{A schematic diagram of Quantum Error Detection (QED).
The projective measurements corresponding to eigenvalues of
stabilizer generators are represented by arrows. For a
non-degenerate QEC code, after the QED, the wavefunction of the
multiqubit system collapses into one of the orthogonal subspaces
each of which is associated with a single error operator. Therefore,
all errors can be unambiguously discriminated. For degenerate codes,
by performing QED the codespace also collapses into a set
orthogonal subspaces. However, each subspace has multiple degeneracies among $%
k $ error operators in a subset of the operator basis, i.e., $\left\{
E_{m}\right\} _{m=1}^{k}\subset \left\{ E_{i}\right\}
_{i=0}^{d^{2}-1}.$ In this case, one cannot distinguish between
different operators within a particular subset $\left\{
E_{m}\right\} _{m=1}^{k_{0}}$.}} \label{Fig-QED} 
\end{figure*}

For a qudit in a general state $\left\vert \phi _{c}\right\rangle $
in the code space, and for arbitrary error basis elements $E_{m}$
and $E_{n}$, the
Knill-Laflamme QEC condition for degenerate codes is $%
\left\langle \phi _{c}\right\vert E_{n}^{\dagger }E_{m}\left\vert
\phi _{c}\right\rangle =\alpha _{nm}$, where $\alpha _{nm}$ is a
Hermitian matrix of complex numbers \cite{nielsen-book}. For
nondegenerate codes, the QEC condition reduces to $\left\langle \phi
_{c}\right\vert E_{n}^{\dagger }E_{m}\left\vert \phi _{c}\right\rangle =$ $%
\delta _{nm}$; i.e., in this case the errors always take the code
space to orthogonal subspaces. The difference between nondegenerate
and degenerate codes is illustrated in Fig.~\ref{Fig-QED}. In this
work, we concentrate on a large class of error-correcting codes
known as stabilizer codes \cite{gottesman-thesis}; however, in contrast to QEC, we restrict
our attention almost
entirely to degenerate stabilizer codes as the initial states.
Moreover, by definition of our problem, the recovery/correction step
is not needed or used in our analysis.

A stabilizer code is a subspace $\mathcal{H}_{C}$ of the Hilbert
space of $n$ qubits that is an eigenspace of a given Abelian subgroup $\mathcal{%
S}$ of the $n$-qubit Pauli group with the eigenvalue $+1$
\cite{nielsen-book,gottesman-thesis}. In other words, for $\left\vert
\phi_{c}\right\rangle \in \mathcal{H}_{C}$ and $S_{i}\in \mathcal{S}$, we have $%
S_{i}\left\vert \phi _{c}\right\rangle =\left\vert \phi _{c}\right\rangle $%
, where $S_{i}$'s are the stabilizer \textit{generators} and
$[S_{i},S_{j}]=0$. Consider the action of an arbitrary error
operator $E$ on the stabilizer code $\left\vert \phi
_{c}\right\rangle $, $E\left\vert \phi _{c}\right\rangle $. The
detection of such an error will be possible if the error
operator anticommutes with (at least one of) the stabilizer
generators, $\{S_{i},E\}=0$. I.e, by measuring all generators of the
stabilizer and obtaining one or more negative eigenvalues we can
determine the nature of the error unambiguously as:
\begin{equation*}
S_{i}(E\left\vert \phi _{c}\right\rangle )=-E(S_{i}\left\vert \phi
_{c}\right\rangle )=-(E\left\vert \phi _{c}\right\rangle ).
\end{equation*}

A stabilizer code $[n$,$k$,$d_{c}]$ represents an encoding of $k$
logical qudits into $n$ physical qudits with code distance $d_{c}$,
such that an arbitrary error on any subset of $t=(d_{c}-1)/2$ or
fewer qudits can be
detected by QED measurements. A stabilizer group with $n-k$ generators
has $d^{n-k}$ elements and the code space is $d^{k}$-dimensional. Note
that this is valid when $d$ is a power of a prime \cite{Gottessmanhighd}. The unitary operators that preserve the
stabilizer group by conjugation, i.e., $USU^{\dagger }=S$, are
called the normalizer of the stabilizer group, $N(S)$. Since the
normalizer elements preserve the code space they can be used to
perform certain logical operations in the code space. However, they are
insufficient for performing arbitrary quantum operations
\cite{nielsen-book}.

Similarly to the case of a qubit \cite{mohseni-lidar06}, the DCQD
algorithm for the case of a qudit system consists of two procedures:
(i) a single experimental configuration for characterization of the
quantum dynamical populations, and (ii) $d^{2}-1$ experimental
configurations for characterization of the quantum dynamical
coherences. In both procedures we always use two physical qudits for
the encoding, the principal system $A$ and the ancilla $B$, i.e.,
$n=2$. In procedure (i) -- characterizing the diagonal elements of the
superoperator -- the stabilizer group has two generators. Therefore it has $d^{2}$ elements
and the code space consists of a \textit{single} quantum state (i.e.,
$k=0$). In procedure (ii) -- characterizing the off-diagonal elements of
the
superoperator -- the stabilizer group has a single generator, thus it
has $d$ elements, and the code space is two-dimensional. That is, we
effectively encode a logical qudit (i.e., $k=1$) into two physical
qudits. In next sections, we develop the procedures (i) and (ii) in
detail for a single qudit with $d$ being a prime, and in the appendix
\ref{generalization-Algorithm} we address the generalization to
systems with $d$ being an arbitrary power of a prime.

\section{Characterization of Quantum Dynamical Population}

\label{Population-qudit}

\begin{figure*}[tp]
\begin{center}
\includegraphics[width=13cm,height=2.7cm]{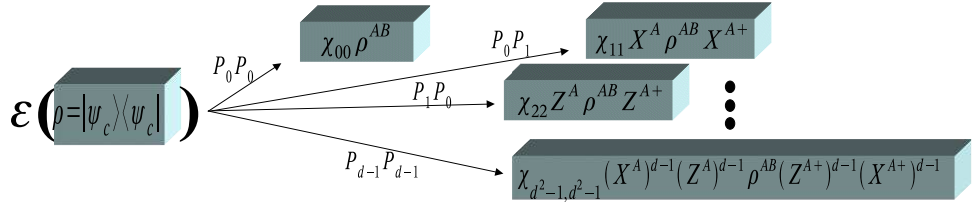}
\end{center}
\caption[A diagram of the error-detection measurement for estimating
quantum dynamical population of a single qudit]{\small{A diagram of
the error-detection measurement for estimating quantum dynamical
population. The arrows represent the projection operators
$P_{k}P_{k^{\prime }}$ corresponding to
different eigenvalues of the two stabilizer generators $S$ and $S^{\prime }$%
. These projective measurements result in a projection of the
wavefunction of the two-qudit systems, after experiencing the
dynamical map, into one of the orthogonal subspaces each of which is
associated to a specific error operator basis.
By calculating the joint probability distribution of all possible outcomes, $%
P_{k}P_{k^{\prime }}$, for $k,k^{\prime }=0,\ldots,d$, we obtain all
$d^{2}$ diagonal elements of the superoperator in a single ensemble
measurement.}} \label{Fig-Pppulation-qudit} 
\end{figure*}

To characterize the diagonal elements of the superoperator, or the
population of the unitary error basis, we use a non-degenerate
stabilizer code. We prepare the principal qudit, $A$, and an ancilla
qudit, $B$, in a common $+1$ eigenstate $\left\vert \phi
_{c}\right\rangle $ of the two unitary operators
$E_{i}^{A}E_{j}^{B}$ and $E_{i^{\prime }}^{A}E_{j^{\prime }}^{B},$
such that $[E_{i}^{A}E_{j}^{B}$,$E_{i^{\prime }}^{A}E_{j^{\prime
}}^{B}]=0$ (e.g. $X^{A}X^{B}$ and $Z^{A}(Z^{B})^{d-1}$). Therefore,
simultaneous measurement of these stabilizer generators at the end
of the dynamical process reveals arbitrary single qudit errors on
the system $A$. The possible outcomes depend on whether a specific
operator in the operator-sum representation of the quantum dynamics
commutes with $E_{i}^{A}E_{j}^{B}$ and $E_{i^{\prime
}}^{A}E_{j^{\prime }}^{B}$, with the eigenvalue $+1$, or with one of
the eigenvalues $\omega , \omega ^{2},\ldots,\omega^{d-1}$. The
projection operators
corresponding to outcomes $\omega ^{k}$ and $\omega ^{k^{\prime }}$, where $k$,$%
k^{\prime }=0,1,\ldots,d-1$, have the form $P_{k}=\frac{1}{d}%
\sum_{l=0}^{d-1}\omega ^{-lk}(E_{i}^{A}E_{j}^{B})^{l}$ and $P_{k^{\prime }}=%
\frac{1}{d}\sum_{l^{\prime }=0}^{d-1}\omega ^{-l^{\prime }k^{\prime
}}(E_{i^{\prime }}^{A}E_{j^{\prime }}^{B})^{l^{\prime }}$. The joint
probability distribution of the commuting Hermitian operators $P_{k}$ and $%
P_{k^{\prime }}$ on the output state $\mathcal{E}(\rho
)=\sum_{m,n}\chi _{mn}~E_{m}\rho E_{n}^{\dagger },$ where $\rho
=\left\vert \phi _{c}\right\rangle \left\langle \phi _{c}\right\vert
,$ is:
\begin{widetext}
\begin{equation*}
\mathrm{Tr}[P_{k}P_{k^{\prime }}\mathcal{E}(\rho )]=\frac{1}{d^{2}}%
\sum\limits_{m,n=0}^{d^{2}-1}\chi
_{mn}\sum_{l=0}^{d-1}\sum_{l^{\prime
}=0}^{d-1}\omega ^{-lk}\omega ^{-l^{\prime }k^{\prime }}\mathrm{Tr}%
[~E_{n}^{\dagger }(E_{i}^{A})^{l}(E_{i^{\prime }}^{A})^{l^{\prime
}}E_{m}(E_{j}^{B})^{l}(E_{j^{\prime }}^{B})^{l^{\prime }}\rho ].
\end{equation*}%
Using $E_{i}E_{m}=\omega ^{i_{m}}E_{m}E_{i}$ and the relation $%
(E_{i}^{A}E_{j}^{B})^{l}(E_{i^{\prime }}^{A}E_{j^{\prime
}}^{B})^{l^{\prime }}\rho =\rho ,$ we obtain:
\begin{equation*}
\mathrm{Tr}[P_{k}P_{k^{\prime }}\mathcal{E}(\rho )]=\frac{1}{d^{2}}%
\sum\limits_{m,n=0}^{d^{2}-1}\chi
_{mn}\sum_{l=0}^{d-1}\sum_{l^{\prime }=0}^{d-1}\omega
^{(i_{m}-k)l}\omega ^{(i_{m}^{\prime }-k^{\prime })l^{\prime
}}\delta _{mn},
\end{equation*}%
\end{widetext}
where we have used the QED condition for nondegenerate codes:
\begin{equation*}
\mathrm{Tr}[E_{n}^{\dagger }E_{m}\rho ]=\left\langle \phi
_{c}\right\vert E_{n}^{\dagger }E_{m}\left\vert \phi
_{c}\right\rangle =\delta _{mn},
\end{equation*}%
i.e., the fact that different errors should take the code space to
orthogonal subspaces, in order for errors to be unambiguously
detectable, see Fig.~\ref{Fig-Pppulation-qudit}. Now, using the
discrete Fourier transform identities $\sum_{l=0}^{d-1}\omega
^{(i_{m}-k)l}=d\delta _{i_{m},k} $ and $\sum_{l^{\prime
}=0}^{d-1}\omega ^{(i_{m}^{\prime }-k^{\prime })l^{\prime }}=d\delta
_{i_{m}^{\prime },k^{\prime }}$, we obtain:
\begin{equation}
\mathrm{Tr}[P_{k}P_{k^{\prime }}\mathcal{E}(\rho
)]=\sum\limits_{m=0}^{d^{2}-1}\chi _{mm}~\delta _{i_{m},k}\delta
_{i_{m}^{\prime },k^{\prime }}=\chi _{m_{0}m_{0}}.~
\label{DiagonalEq}
\end{equation}%
Here, $m_{0}$ is defined through the relations $i_{m_{0}}=k$ and $%
i_{m_{0}}^{\prime }=k^{\prime }$, i.e., $E_{m_{0}}$ is the unique
error operator that anticommutes with the stabilizer operators with
a fixed pair of eigenvalues $\omega ^{k}$ and $\omega ^{k^{\prime
}}$ corresponding to
the experimental outcomes $k$ and $k^{\prime }$. Since each $P_{k}$ and $%
P_{k^{\prime }}$ operator has $d$ eigenvalues, we have $d^{2}$
possible outcomes, which gives us $d^{2}$ linearly independent
equations. Therefore, we can \emph{characterize all the diagonal
elements of the superoperator with a single ensemble measurement}
and $2d$ detectors.

In order to investigate the properties of the pure state $\left\vert
\phi _{c}\right\rangle $, we note that the code space is
one-dimensional (i.e., it has only one vector) and can be Schmidt
decomposed as $\left\vert \phi _{c}\right\rangle
=\sum_{k=0}^{d-1}\lambda _{k}\left\vert k\right\rangle
_{A}\left\vert k\right\rangle _{B}$, where $\lambda _{k}$ are
non-negative real numbers. Suppose $Z\left\vert k\right\rangle
=\omega ^{k}\left\vert k\right\rangle $; without loss of generality
the two stabilizer generators of $\left\vert \phi
_{c}\right\rangle $ can be chosen to be $(X^{A}X^{B})^{q}$ and $%
[Z^{A}(Z^{B})^{d-1}]^{p}$. We then have $\left\langle \phi
_{c}\right\vert (X^{A}X^{B})^{q}\left\vert \phi _{c}\right\rangle
=1$ and $\left\langle \phi _{c}\right\vert
[Z^{A}(Z^{B})^{d-1}]^{p}\left\vert \phi _{c}\right\rangle =1$ for
any $q$ and $p$, where $0\leq q,p<d$. This results in the set of
equations $\sum_{k=0}^{d-1}\lambda _{k}\lambda _{k+q}=1$ for all
$q$, which have only one positive real solution: $\lambda
_{0}=\lambda _{1}=\ldots =\lambda _{k}=1/\sqrt{d}$; i.e., the
stabilizer state, $\left\vert \phi _{c}\right\rangle $, is a
\textit{maximally entangled state} in the Hilbert space of the two
qudits.

In the remaining parts of this paper, we first develop an algorithm
for extracting optimal information about the dynamical coherence of
a $d$-level quantum system (with $d$ being a prime), through a
single experimental configuration, in Sec.~\ref{basic-lemmas}, \ref{coherence-qudit} and %
\ref{linearindep-optimality}. Then, we further develop the algorithm
to obtain complete information about the off-diagonal elements of
the superoperator by repeating the same scheme for different input states, Sec.~\ref%
{rep-algorithm}. In Sec.~\ref{generalization-Algorithm}, we address
the generalization of the DCQD algorithm for qudit systems with $d$
being a power of a prime. In the first step, in the next section, we
establish the required notation by introducing some lemmas and
definitions.

\section{Basic lemmas and definitions}

\label{basic-lemmas}

\begin{lemma}
\label{unique} Let $0\leq q,p,q^{\prime },p^{\prime }<d$, where $d$
is prime. Then, for given $q$, $p$, $q^{\prime}$ and
$k(\mathrm{mod}~d)$,
there is a unique $p^{\prime }$ that solves $pq^{\prime }-qp^{\prime }=k~(%
\mathrm{mod}~d).$
\end{lemma}
\begin{proof}
We have $pq^{\prime }-qp^{\prime }=k~(\mathrm{mod}~d)=k+td,$ where
$t$ is an
integer. The possible solutions for $p^{\prime }$ are indexed by $t$ as $%
p^{\prime }(t)=(pq^{\prime }-k-td)/q.$ We now show that if
$p^{\prime }(t_{1})$ is a solution for a specific value $t_{1}$,
there exists no other integer $t_{2}\neq t_{1}$ such that $p^{\prime
}(t_{2})$ is another independent solution to this equation,$~$i.e.,
$p^{\prime }(t_{2})\neq p^{\prime }(t_{1})(\mathrm{mod}~d).$ First,
note that if $p^{\prime }(t_{2})$ is another solution then we have
$p^{\prime }(t_{1})=p^{\prime }(t_{2})+(t_{2}-t_{1})d/q.$ Since $d$
is prime, there are two possibilities: a) $q$ divides
$(t_{2}-t_{1}),$ then $(t_{2}-t_{1})d/q=\pm nd,$ where $n$ is
a positive integer; therefore we have $p^{\prime }(t_{2})=p^{\prime }(t_{1})(%
\mathrm{mod}~d),\,\ $which contradicts our assumption that
$p^{\prime }(t_{2})$ is an independent solution from $p^{\prime
}(t_{1})$. b) $q$ does not divide $(t_{2}-t_{1}),$ then
$(t_{2}-t_{1})d/q$ is not a integer, which
is unacceptable. Thus, we have $t_{2}=t_{1},$ i.e., the solution $%
p^{\prime }(t)$ is unique.

Note that the above argument does not hold if $d$ is not prime, and
therefore, for some $q^{\prime }$ there could be more than one
$p^{\prime }$ that satisfies $pq^{\prime }-qp^{\prime }\equiv
k~(\mathrm{mod}~d)$. In general, the validity of this lemma relies
on the fact that $\mathtt{ \mathbb{Z} }_{d}$ is a field only for
prime $d$.
\end{proof}

\begin{lemma}
\label{subsetsW} For any unitary error operator basis $E_{i}$ acting
on a
Hilbert space of dimension $d$, where $d$ is a prime and $i=0,1,\ldots,d^{2}-1$%
, there are $d$ unitary error operator basis elements, $E_{j}$, that
anticommute with $E_{i}$ with a specific eigenvalue $\omega ^{k}$, i.e., $%
E_{i}E_{j}=\omega ^{k}E_{j}E_{i}$, where $k=0,\ldots,d-1$.
\end{lemma}
\begin{proof}
We have $E_{i}E_{j}=\omega ^{pq^{\prime }-qp^{\prime }}E_{j}E_{i}$, where $%
0\leq q$,$p$,$q^{\prime }$,$p^{\prime }<d$, and $pq^{\prime
}-qp^{\prime
}\equiv k~(\mathrm{mod}~d)$. Therefore, for fixed $q$, $p$, and $k$ $(\mathrm{mod}%
~d) $ we need to show that there are $d$ solutions $(q^{\prime
}$,$p^{\prime })$. According to Lemma~\ref{unique}, for any
$q^{\prime }$ there is only
one $p^{\prime }$ that satisfies $pq^{\prime }-qp^{\prime }=k~(\mathrm{mod}%
~d);$ but $q^{\prime }$can have $d$ possible values, therefore there
are $d$ possible pairs of $(q^{\prime }$,$p^{\prime })$.
\end{proof}
\begin{definition}
\label{definition Wk} We introduce $d$ different subsets, $W_{k}^{i}$, $%
k=0,1,\ldots,d-1,$ of a unitary error operator basis $\{E_{j}\}$ (i.e. $%
W_{k}^{i}\subset \{E_{j}\})$. Each subset contains $d$ members which
all anticommute with a particular basis element $E_{i},$ where
$i=0,1,\ldots,d^{2}-1,$ with fixed eigenvalue $\omega ^{k}.$ The
subset $W_{0}^{i}$ which includes $E_{0}$ and $E_{i}$ is in fact an
Abelian subgroup of the single-qudit Pauli group, $G_{1}$.
\end{definition}

\section{Characterization of Quantum Dynamical Coherence}

\label{coherence-qudit}

For characterization of the coherence in a quantum dynamical process
acting on a qudit system, we prepare a two-qudit quantum system in a
non-separable
eigenstate $\left\vert \phi _{ij}\right\rangle $ of a unitary operator $%
S_{ij}=E_{i}^{A}E_{j}^{B}$. We then subject the qudit $A$ to the
unknown dynamical map, and measure the sole stabilizer operator
$S_{ij}$ at the output state. Here, the state\ $\left\vert \phi
_{ij}\right\rangle $ is in fact a degenerate code space, since all
the operators $E_{m}^{A}$ that anticommute with $E_{i}^{A},$ with a
particular eigenvalue $\omega ^{k}$, perform the same transformation
on the code space and cannot be distinguished by the stabilizer
measurement. If we express the spectral
decomposition of $S_{ij}=E_{i}^{A}E_{j}^{B}$ as $S_{ij}=\sum_{k}$ $%
\omega ^{k}P_{k}$, the projection operator corresponding to the
outcome $\omega ^{k}$ can be written as
$P_{k}=\frac{1}{d}\sum_{l=0}^{d-1}\omega
^{-lk}(E_{i}^{A}E_{j}^{B})^{l}$. The post-measurement state of the
system, up a normalization factor, will be:
\begin{widetext}
\begin{equation*}
P_{k}\mathcal{E}(\rho )P_{k}=\frac{1}{d^{2}}\sum\limits_{m,n=0}^{d^{2}-1}%
\chi _{mn}\sum_{l=0}^{d-1}\sum_{l^{\prime }=0}^{d-1}\omega
^{-lk}\omega ^{l^{\prime }k}[(E_{i}^{A}E_{j}^{B})^{l}E_{m}\rho
E_{n}^{\dagger }(E_{i}^{A\dagger }E_{j}^{B\dagger })^{l^{\prime }}].
\end{equation*}%
Using the relations $E_{i}E_{m}=\omega ^{i_{m}}E_{m}E_{i}$,
$E_{n}^{\dagger
}E_{i}^{\dagger }=\omega ^{-i_{n}}E_{i}^{\dagger }E_{n}^{\dagger }$ and $%
(E_{i}^{A}E_{j}^{B})^{l}\rho (E_{i}^{A\dagger }E_{j}^{B\dagger
})^{l^{\prime }}=\rho $ we have:
\begin{equation*}
P_{k}\mathcal{E}(\rho )P_{k}=\frac{1}{d^{2}}\sum_{l=0}^{d-1}\omega
^{(i_{m}-k)l}\sum_{l^{\prime }=0}^{d-1}\omega ^{(k-i_{n})l^{\prime
}}\sum\limits_{m,n=0}^{d^{2}-1}\chi _{mn}E_{m}\rho E_{n}^{\dagger }.
\end{equation*}%
Now, using the discrete Fourier transform properties
$\sum_{l=0}^{d-1}\omega ^{(i_{m}-k)l}=d\delta _{i_{m},k}$ and
$\sum_{l^{\prime }=0}^{d-1}\omega ^{(k-i_{n})l^{\prime }}=d\delta
_{i_{n},k}\,,$\ we obtain:
\begin{equation}
P_{k}\mathcal{E}(\rho )P_{k}=\sum\limits_{m}\chi _{mm}~E_{m}^{A}\rho
E_{m}^{A\dagger }+\sum_{m<n}(\chi _{mn}~E_{m}^{A}\rho
E_{n}^{A\dagger }+\chi _{mn}^{*}~E_{n}^{A}\rho E_{m}^{A\dagger }).
\label{PostMeasurmentStates}
\end{equation}%
\end{widetext}
Here, the summation runs over all $E_{m}^{A}$ and $E_{n}^{B}$ that
belong to the same $W_{k}^{i}$; see Lemma~\ref{subsetsW}. I.e., the
summation is over all unitary operator basis elements $E_{m}^{A}$
and $E_{n}^{B}$ that anticommute with $E_{i}^{A}$ with a particular
eigenvalue $\omega ^{k}$. Since the number of elements in each
$W_{k}$ is $d$, the state of the two-qudit system after the
projective measurement comprises $d+2[d(d-1)/2]=d^{2}$ terms. The
probability of getting the outcome $\omega ^{k}$ is:
\begin{equation}
\mathrm{Tr}[P_{k}\mathcal{E}(\rho )]=\sum_{m}\chi _{mm}+2\sum_{m<n}\mathrm{Re%
}[\chi _{mn}~\mathrm{Tr}(E_{n}^{A}{}^{\dagger }E_{m}^{A}\rho )].
\label{StabilizerProbEq}
\end{equation}%
Therefore, the normalized post-measurement states are $\rho _{k}=P_{k}%
\mathcal{E}(\rho )P_{k}/$\textrm{Tr}$[P_{k}\mathcal{E}(\rho )]$.
These $d$ equations provide us with information about off-diagonal
elements of the superoperator iff \textrm{Tr}$[(E_{n}^{A})^{\dagger
}E_{m}^{A}\rho ]\neq 0$. Later we will derive some general
properties of the state $\rho $ such that this condition can be
satisfied.

\begin{center}
\begin{figure*}[tp]
\begin{center}
\includegraphics[width=11cm,height=2.6cm]{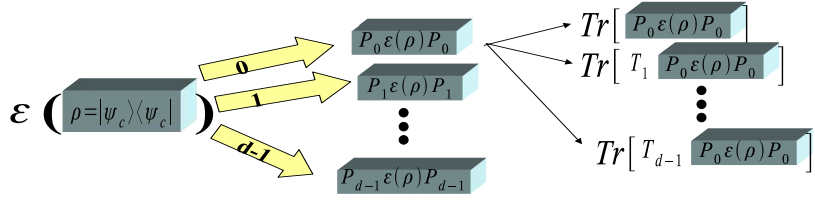}
\end{center}
\caption[A diagram of the error-detection measurement for estimating
quantum dynamical coherence of a single qudit]{\small{A diagram of
the error-detection measurement for estimating quantum dynamical
coherence: we measure the sole stabilizer generator at the output
state, by applying projection operators corresponding to its
different eigenvalues $P_{k}$. We also measure $d-1$ commuting
operators that belong to the normalizer group. Finally, we calculate
the probability of each stabilizer outcome, and joint probability
distributions of the normalizers and the stabilizer outcomes.
Optimally, we can obtain $d^{2}$ linearly independent equations by
appropriate selection of the normalizer operators as it is shown in
the next section.}} \label{Fig-coherence-qudit} 
\end{figure*}
\end{center}

Next we measure the expectation value of any other unitary operator
basis
element $T_{rs}=E_{r}^{A}E_{s}^{B}$ on the output state, such that $%
E_{r}^{A}\neq I$, $E_{s}^{B}\neq I$,$~T_{rs}\in N(S)$ and
$T_{rs}\neq (S_{ij})^{a}$, where $0\leq a<d$. Let us write the
spectral decomposition of $T_{rs}$ as $T_{rs}=\sum\limits_{k^{\prime
}}$ $\omega ^{k^{\prime }}P_{k^{\prime }}$. The joint probability
distribution of the commuting
Hermitian operators $P_{k}$ and $P_{k^{\prime }}$ on the output state $%
\mathcal{E}(\rho )$ is $\mathrm{Tr}[P_{k^{\prime }}P_{k}\mathcal{E}(\rho )]$%
. The average of these joint probability distributions of $P_{k}$ and $%
P_{k^{\prime }}$ over different values of $k^{\prime }$ becomes: $%
\sum_{k^{\prime }}\omega ^{k^{\prime }}\mathrm{Tr}[ P_{k^{\prime }}P_{k}%
\mathcal{E}(\rho )]=\mathrm{Tr}[T_{rs}P_{k}\mathcal{E}(\rho )]=\mathrm{Tr}%
(T_{rs}\rho _{k})$, which can be explicitly written as:
\begin{widetext}
\begin{eqnarray*}
\mathrm{Tr}(T_{rs}\rho _{k}) &=& \sum_{m}\chi _{mm}~\mathrm{Tr}%
(E_{m}^{A}{}^{\dagger }E_{r}^{A}E_{s}^{B}E_{m}^{A}\rho )  
+\sum_{m<n}[\chi _{mn}~\mathrm{Tr}(E_{n}^{A}{}^{\dagger
}E_{r}^{A}E_{s}^{B}E_{m}^{A}\rho )+\chi _{mn}^{*}~\mathrm{Tr}%
(E_{m}^{A}{}^{\dagger }E_{r}^{A}E_{s}^{B}E_{n}^{A}\rho )].
\end{eqnarray*}
Using $E_{r}^{A}E_{m}^{A}=\omega ^{r_{m}}E_{m}^{A}E_{r}^{A}$ and $%
E_{r}^{A}E_{n}^{A}=\omega ^{r_{n}}E_{n}^{A}E_{r}^{A}$ this becomes:
\begin{eqnarray}
\hskip -2.5mm\mathrm{Tr}(T_{rs}\rho _{k}) &=&\frac{1}{\mathrm{Tr}[P_{k}\mathcal{E}(\rho )]%
}\left(\sum_{m}\omega ^{r_{m}}\chi _{mm}~\mathrm{Tr}(T_{rs}\rho )  
+\sum_{m<n}\left[ \omega ^{r_{m}}\chi _{mn}~\mathrm{Tr}(E_{n}^{A}{}^{%
\dagger }E_{m}^{A}T_{rs}\rho )+\omega ^{r_{n}}\chi _{mn}^{* }~\mathrm{Tr}%
(E_{m}^{A}{}^{\dagger }E_{n}^{A}T_{rs}\rho )\right] \right).
\label{NormalizerEq}
\end{eqnarray}%
\end{widetext}
Therefore, we have an additional set of $d$ equations
to identify the off-diagonal elements of the superoperator, provided that \textrm{Tr}$%
(E_{n}^{A}{}^{\dagger }E_{m}^{A}T_{rs}\rho )\neq 0$. Suppose we now
measure another unitary operator $T_{r^{\prime }s^{\prime
}}=E_{r^{\prime }}^{A}E_{s^{\prime }}^{B}$ that commutes with
$S_{ij}$, i.e. $T_{r^{\prime }s^{\prime }}\in N(S)$, and also
commutes with $T_{rs},$ and satisfies the
relations $T_{r^{\prime }s^{\prime }}\neq T_{rs}^{b}S_{ij}{}^{a}$ (where $%
0\leq a$,$b<d$), $E_{r}^{A}\neq I$ and $E_{s}^{B}\neq I$. Such a
measurement results in $d$ equations for $\mathrm{Tr}(T_{r^{\prime
}s^{\prime }}\rho _{k})$, similar to those for
$\mathrm{Tr}(T_{rs}\rho _{k})$. However, for these equations to be
useful for characterization of the dynamics, one needs show that
they are all linearly independent. In the next section, we find the
maximum number of independent and commutating unitary operators
$T_{rs}$
such that their expectation values on the output state, $\mathrm{Tr}%
(T_{rs}\rho _{k})$, result in linearly independent equations to be
$d-1$, see Fig.~\ref{Fig-coherence-qudit}. I.e., we find an optimal
Abelian set of unitary operators such that the joint probability
distribution functions of their eigenvalues and stabilizer
eigenvalues at the output state are linearly independent.

\section{Linear independence and optimality of measurements}

\label{linearindep-optimality}

Before presenting the proof of linear independence of the functions $\mathrm{%
Tr}(T_{rs}\rho _{k})$ and of the optimality of the DCQD algorithm,
we need to introduce the following lemmas and definitions.

\begin{lemma}
\label{order of N} If a stabilizer group, $S$, has a single
generator, the order of its normalizer group, $N(S)$, is $d^{3}$.
\end{lemma}
\begin{proof}
Let us consider the sole stabilizer generator
$S_{12}=E_{1}^{A}E_{2}^{B}$, and a typical normalizer element
$T_{1^{\prime }2^{\prime }}=E_{1^{\prime
}}^{A}E_{2^{\prime }}^{B}$, where $E_{1}^{A}=X^{q_{1}}Z^{p_{1}}$,$%
~E_{2}^{B}=X^{q_{2}}Z^{p_{2}}$,$~E_{1^{\prime }}^{A}=X^{q_{1^{\prime
}}}Z^{p_{1^{\prime }}}$ and $E_{2^{\prime }}^{B}=X^{q_{2^{\prime
}}}Z^{p_{2^{\prime }}}$. Since $S_{12}$ and $T_{1^{\prime }2^{\prime
}}$ commute, we have $S_{12}T_{1^{\prime }2^{\prime }}=\omega
^{\sum_{i=1}^{2}p_{i}q_{i^{\prime }}^{\prime }-q_{i}p_{i^{\prime
}}^{\prime }}T_{1^{\prime }2^{\prime }}S_{12}$, where
$\sum_{i=1}^{2}p_{i}q_{i^{\prime }}^{\prime }-q_{i}p_{i^{\prime
}}^{\prime }\equiv 0~(\mathrm{mod}~d)$. We
note that for any particular code with a single stabilizer generator, all $%
q_{1}$,$p_{1}$,$q_{2}$ and $p_{2}$ are fixed. Now, by Lemma
\ref{unique}, for given values of $q_{1}^{\prime }$,$p_{1}^{\prime
}$ and $q_{2}^{\prime }$ there is only one value for $p_{2}^{\prime
}$ that satisfies the above
equation. However, each of $q_{1}^{\prime }$,$p_{1}^{\prime }$ and $%
q_{2}^{\prime }$ can have $d$ different values. Therefore, there are
$d^{3}$ different normalizer elements $T_{1^{\prime }2^{\prime }}$.
\end{proof}

\begin{lemma}
\label{order of A} Each Abelian subgroup of a normalizer, which
includes the
stabilizer group $\{S_{ij}^{a}{}\}$ as a proper subgroup, has order $d^{2}$.$%
\ \ $
\end{lemma}
\begin{proof}
Suppose $T_{rs}$ is an element of $N(S)$, i.e., it commutes with
$S_{ij}$.
Moreover, all unitary operators of the form $T_{rs}^{b}S_{ij}{}^{a}$, where $%
0\leq a$,$b<d$, also commute. Therefore, any Abelian subgroup of the
normalizer, $A\subset N(S)$, which includes $\{S_{ij}^{a}{}\}$ as a
proper subgroup, is at least order of $d^{2}$. Now let $T_{r^{\prime
}s^{\prime }}$ be any other normalizer element, i.e., $T_{r^{\prime
}s^{\prime }}\neq T_{rs}^{b}S_{ij}^{a}$ with $0\leq a$,$b<d$, which
belongs to the same Abelian subgroup $A$. In this case, any operator
of the form $T_{r^{\prime }s^{\prime }}^{b^{\prime
}}T_{rs}^{b}S_{ij}^{a}$ would also belong to $A$. Then all elements
of the normalizer should commute or $A=N(S)$, which is unacceptable.
Thus, either $T_{r^{\prime }s^{\prime }}=T_{rs}^{b}S_{ij}^{a}$ or
$T_{r^{\prime }s^{\prime }}\notin A$, i.e., the order of the Abelian
subgroup $A$ is at most $d^{2}$.
\end{proof}

\begin{lemma}
\label{d+1 of A} There are $d+1$ Abelian subgroups, $A$, in the normalizer $%
N(S)$.
\end{lemma}
\begin{proof}
Suppose that the number of Abelian subgroups which includes the
stabilizer group as a proper subgroup is $n$. Using Lemmas
\ref{order of N} and \ref{order of A}, we have:
$d^{3}=nd^{2}-(n-1)d$, where the term $(n-1)d$ has been subtracted
from the total number of elements of the normalizer due to the fact
that the elements of the stabilizer group are common to all Abelian
subgroups. Solving this equation for $n$, we find that $n=\frac{d^{2}-1}{d-1}%
=d+1$.
\end{proof}

\begin{lemma}
\label{MUA} The basis of eigenvectors defined by $d+1$ Abelian subgroups of $%
N(S)$ are mutually unbiased.
\end{lemma}
\begin{proof}
It has been shown \cite{som-mub} that if a set of $d^{2}-1$
traceless and
mutually orthogonal $d\times d$ unitary matrices can be partitioned into $%
d+1 $ subsets of equal size, such that the $d-1$ unitary operators
in each subset commute, then the basis of eigenvectors corresponding
to these
subsets are mutually unbiased. We note that, based on Lemmas \ref{order of N}%
, \ref{order of A} and \ref{d+1 of A}, and in the code space (i.e.,
up to multiplication by the stabilizer elements $\{S_{ij}^{a}{}\}$),
the normalizer $N(S)$ has $d^{2}-1$ nontrivial elements, and each
Abelian subgroup $A$, has $d-1$ nontrivial commuting operators.
Thus, the bases of eigenvectors defined by $d+1$ Abelian subgroups
of $N(S)$ are mutually unbiased.
\end{proof}

\begin{lemma}
\label{d cosets} Let $C$ be a cyclic subgroup of $A$, i.e.,
$C\subset A\subset N(S)$. Then, for any fixed $T\in A$, the number
of distinct left (right) cosets, $TC~(CT)$, in each $A$ is $d$.
\end{lemma}
\begin{proof}
We note that the order of any cyclic subgroup $C\subset A$, such as $%
T_{rs}^{b}$ with $0\leq b<d$, is $d$. Therefore, by Lemma \ref{order
of A}, the number of distinct cosets in each $A$ is
$\frac{d^{2}}{d}=d$.
\end{proof}

\begin{definition}
We denote the cosets of an (invariant) cyclic subgroup, $C_{a}$, of
an Abelian subgroup of the normalizer, $A_{v}$, by $A_{v}/C_{a}$,
where $v=1,2,\ldots,d+1$. We also represent generic members of $A_{v}/C_{a}$ as $%
T_{rs}^{b}S_{ij}^{a},$ where $0\leq a$,$b<d$. The members of a
specific coset $A_{v}/C_{a_{0}}$ are denoted as
$T_{rs}^{b}S_{ij}^{a_{0}}$, where $a_{0}$ represents a fixed power
of stabilizer generator $S_{ij},$ that labels a particular coset
$A_{v}/C_{a_{0}}$, and $b$ $(0\leq b<d)$ labels different members of
that particular coset.
\end{definition}

\begin{lemma}
\label{diff eigenvalues for Er and Ei}The elements of a coset, $%
T_{rs}^{b}S_{ij}^{a_{0}}$ (where $T_{rs}=E_{r}^{A}E_{s}^{B}$, $%
S_{ij}=E_{i}^{A}E_{j}^{B}$ and $0\leq b<d)$ anticommute with
$E_{i}^{A}$ with different eigenvalues $\omega ^{k}$. I.e., there
are no two different members of a coset, $A_{v}/C_{a_{0}}$, that
anticommute with $E_{i}^{A}$ with the same eigenvalue.
\end{lemma}
\begin{proof}
First we note that for each
$T_{rs}^{b}=(E_{r}^{A})^{b}(E_{s}^{B})^{b}$, the unitary operators
acting only on the principal subsystem, $(E_{r}^{A})^{b}$,
must satisfy either (a) $(E_{r}^{A})^{b}=E_{i}^{A}$ or (b) $%
(E_{r}^{A})^{b}\neq E_{i}^{A}$. In the case (a), and due to $[T_{rs}$,$S_{ij}]=0$%
, we should also have $(E_{s}^{B})^{b}=E_{j}^{B},$ which results in $%
T_{rs}^{b}=S_{ij};$ i.e., $T_{rs}^{b}$ is a stabilizer and not a
normalizer. This is unacceptable. In the case (b), in particular for $b=1$, we have $%
E_{r}^{A}E_{i}^{A}=\omega ^{r_{i}}E_{i}^{A}E_{r}^{A}$. Therefore,
for arbitrary $b$ we have $(E_{r}^{A})^{b}E_{i}^{A}=\omega
^{br_{i}}E_{i}^{A}(E_{r}^{A})^{b}$. Since $0\leq b<d$, we conclude that $%
\omega ^{br_{i}}\neq \omega ^{b^{\prime }r_{i}}$ for any two
different values of $b$ and $b^{\prime }$.

As a consequence of this lemma, different $(E_{r}^{A})^{b}$, for $0\leq b<d$%
, belong to different $W_{k}^{i}$'s.
\end{proof}

\begin{lemma}
\label{diff eigenvalues for Er and Em} For any fixed unitary operator $%
E_{r}^{A}\in W_{k}^{i}$ , where $k\neq 0$, and any other two
independent
operators $E_{m}^{A}$ and $E_{n}^{A}$ that belong to the same $W_{k}^{i}\,$%
,\ we always have $\omega ^{r_{m}}\neq \omega ^{r_{n}},$ where $%
E_{r}^{A}E_{m}^{A}=\omega ^{r_{m}}E_{m}^{A}E_{r}^{A}$ and $%
E_{r}^{A}E_{n}^{A}=\omega ^{r_{n}}E_{n}^{A}E_{r}^{A}$.
\end{lemma}
\begin{proof}
We need to prove for operators $E_{r}^{A}$,$E_{m}^{A}$,$E_{n}^{A}\in $ $%
W_{k}^{i}$ (where $k\neq 0$),$\ $that we always have: $E_{m}^{A}\neq
E_{n}^{A}\Longrightarrow \omega ^{r_{m}}\neq \omega ^{r_{n}}$. Let
us prove
the converse: $\omega ^{r_{m}}=\omega ^{r_{n}}\Longrightarrow $ $%
E_{m}^{A}=E_{n}^{A}$. We define $E_{i}^{A}=X^{q_{i}}Z^{p_{i}}$, $%
E_{r}^{A}=X^{q_{r}}Z^{p_{r}}$, $E_{m}^{A}=X^{q_{m}}Z^{p_{m}}$, $%
E_{n}^{A}=X^{q_{n}}Z^{p_{n}}$. Based on the definition of subsets
$W_{k}^{i}$ with $k\neq 0$, we have: $p_{i}q_{m}-q_{i}p_{m}\equiv
p_{i}q_{n}-q_{i}p_{n}=k~(\mathrm{mod}~d)=k+td$ (I), where $t$ is an
integer number. We need to show if $\ p_{r}q_{m}-q_{r}p_{m}\equiv
p_{r}q_{n}-q_{r}p_{n}=k^{\prime }(\mathrm{mod}~d)=k^{\prime
}+t^{\prime }d$ (II), then $E_{m}^{A}=E_{n}^{A}$.

We divide the equations (I) by $q_{i}q_{m}$ or $q_{i}q_{n}$ to get: $\frac{p_{i}%
}{q_{i}}=\frac{k+td}{q_{i}q_{m}}+\frac{p_{m}}{q_{m}}=\frac{k+td}{q_{i}q_{n}}+%
\frac{p_{n}}{q_{n}}$ (I'). We also divide the equations (II) by $q_{r}q_{m}$ or $%
q_{r}q_{n}$ to get: $\frac{p_{r}}{q_{r}}=\frac{k^{\prime }+t^{\prime }d}{%
q_{r}q_{m}}+\frac{p_{m}}{q_{m}}=\frac{k^{\prime }+t^{\prime }d}{q_{r}q_{n}}+%
\frac{p_{n}}{q_{n}}$ (II'). By subtracting the equation (II') from (I') we get: $%
q_{n}(\frac{k+td}{q_{i}}-\frac{k^{\prime }+t^{\prime }d}{q_{r}})=q_{m}(\frac{%
k+td}{q_{i}}-\frac{k^{\prime }+t^{\prime }d}{q_{r}})$ (1).
Similarly, we can
obtain the equation $p_{n}(\frac{k+td}{p_{i}}-\frac{k^{\prime }+t^{\prime }d%
}{p_{r}})=p_{m}(\frac{k+td}{p_{i}}-\frac{k^{\prime }+t^{\prime
}d}{p_{r}})$ (2). Note that the expressions within the brackets in
both equations (1) or
(2) cannot be simultaneously zero, because it will result in $%
p_{i}q_{r}-q_{i}p_{r}=0$, which is unacceptable for $k\neq 0$.
Therefore, the expression within the brackets in at least one of the
equations (1) or (2) is non-zero. This results in $q_{n}=q_{m}$
and/or $p_{n}=p_{m}$. Consequently, considering the equation (I), we
have $E_{m}^{A}=E_{n}^{A}$.
\end{proof}

\subsection{Linear independence of the joint distribution functions}

\begin{theorem}
\label{theorem LI} The expectation values of normalizer elements on
a post-measurement state, $\rho _{k}$, are linearly independent if
these elements are the $d-1$ nontrivial members of a coset
$A_{v}/C_{a_{0}}$. I.e., for two independent operators $T_{rs}$,
$T_{r^{\prime }s^{\prime }}\in
A_{v}/C_{a_{0}}$, we have $\mathrm{Tr}(T_{rs}\rho _{k})\neq c~\mathrm{Tr}%
(T_{r^{\prime }s^{\prime }}\rho _{k})$, where $c$ is an arbitrary
complex number.
\end{theorem}
\begin{proof}
We know that the elements of a coset can be written as $%
T_{rs}^{b}S_{ij}^{a_{0}}=(E_{r}^{A}E_{s}^{B})^{b}S_{ij}^{a_{0}}$, where $%
b=1,2,\ldots,d-1$.$\ $\ We also proved that $(E_{r}^{A})^{b}$
belongs to different $W_{k}^{i}$ ($k\neq 0)$ for different values of $b$ (see Lemma \ref%
{diff eigenvalues for Er and Ei}). Therefore, according to Lemma
\ref{diff eigenvalues for Er and Em} and regardless of the outcome
of $k$ (after
measuring the stabilizer $S_{ij})$, there exists one member in the coset $%
A_{v}/C_{a_{0}}$ that has different eigenvalues $\omega ^{r_{m}}$
with all
(independent) members $E_{m}^{A}\in W_{k}^{i}$. The expectation value of $%
T_{rs}^{b}S_{ij}^{a_{0}}$ is:
\begin{widetext}
\begin{eqnarray}
\mathrm{Tr}(T_{rs}^{b}S_{ij}^{a_{0}}\rho _{k}) &=& \sum_{m}\chi _{mm}~%
\mathrm{Tr}(E_{m}^{A}{}^{\dagger
}T_{rs}^{b}S_{ij}^{a_{0}}E_{m}^{A}\rho ), 
+\sum_{m<n}[\chi _{mn}~\mathrm{Tr}(E_{n}^{A}{}^{\dagger
}T_{rs}^{b}S_{ij}^{a_{0}}E_{m}^{A}\rho )+\chi _{mn}^{* }~\mathrm{Tr}%
(E_{m}^{A}{}^{\dagger }T_{rs}^{b}S_{ij}^{a_{0}}E_{n}^{A}\rho )],\\
\mathrm{Tr}(T_{rs}^{b}\rho _{k}) &=& \sum_{m}\omega ^{br_{m}}\chi _{mm}~%
\mathrm{Tr}(T_{rs}^{b}\rho )  
+\sum_{m<n}[\omega
^{br_{m}}\chi_{mn}~\mathrm{Tr}(E_{n}^{A}{}^{\dagger
}E_{m}^{A}T_{rs}^{b}\rho )+\omega ^{br_{n}}\chi _{mn}^{* }~\mathrm{Tr}%
(E_{m}^{A}{}^{\dagger }E_{n}^{A}T_{rs}^{b}\rho )],
\end{eqnarray}
\end{widetext}
where $\omega ^{r_{m}}\neq \omega ^{r_{n}}\neq\ldots $ for all elements $%
E_{m}^{A},E_{n}^{A},\ldots$ that belong to a specific $W_{k}^{i}$.
Therefore, for two independent members of a coset denoted by $b$ and
$b^{\prime }$ (i.e., $b\neq $ $b^{\prime })$, we have $(\omega
^{b^{\prime }r_{m}}, \omega ^{b^{\prime }r_{n}},\ldots)\neq
c~(\omega ^{br_{m}},
\omega ^{br_{n}},\ldots)$ for all values of $0\leq b$,$b^{\prime }<d$, and any complex number $%
c$. We also note that we have \textrm{Tr}$(E_{n}^{A}{}^{\dagger
}E_{m}^{A}T_{rs}^{b}\rho )\neq c~\mathrm{Tr}(E_{n}^{A}{}^{\dagger
}E_{m}^{A}T_{rs}^{b^{\prime }}\rho )$, since $T_{rs}^{b^{\prime
}-b}$ is a normalizer, not a stabilizer element, and its action on
the state cannot be expressed as a global phase. Thus, for any two
independent members of a coset $A_{v}/C_{a_{0}}$, we always have
\textrm{Tr}$(T_{rs}^{b^{\prime }}\rho _{k})\neq c$
$\mathrm{Tr}(T_{rs}^{b}\rho _{k})$.
\end{proof}

In summary, after the action of the unknown dynamical process, we
measure the eigenvalues of the stabilizer generator,
$E_{i}^{A}E_{j}^{B}$, that has $d$ eigenvalues for
$k=0,1,\ldots,d-1$ and provides $d$ linearly independent equations
for the real and imaginary parts of $\chi _{mn}$. This is due to
that the outcomes corresponding to different eigenvalues of a
unitary operator are independent. We also measure expectation values
of all the $d-1$ independent and commuting normalizer operators
$T_{rs}^{b}S_{ij}^{a_{0}}\in A_{v}/C_{a_{0}}$, on the
post-measurement state $\rho _{k}$, which provides $(d-1)$ linearly
independent equations for each outcome $k$ of the stabilizer
measurements. Overall, we obtain $d+d(d-1)=d^{2}$ linearly
independent equations for characterization of the real and imaginary
parts of $\chi _{mn}$ by a single ensemble measurement. In the
following, we show that the above algorithm is optimal. I.e.,
there does not exist any other possible strategy that can provide more than $%
\log _{2}d^{2}$ bits of information by a single measurement on the
output state $\mathcal{E}(\rho )$.

\subsection{Optimality}

\begin{theorem}
The maximum number of commuting normalizer elements that can be
measured simultaneously to provide linear independent equations for
the joint
distribution functions $\mathrm{Tr} (T_{rs}^{b} S_{ij}^{a} \rho _{k})$ is $%
d-1$.
\end{theorem}
\begin{proof}
Any Abelian subgroup of the normalizer has order $d^{2}$ (see Lemma \ref%
{order of A}). Therefore, the desired normalizer operators should
all belong to a particular $A_{v}$ and are limited to $d^{2}$
members. We already
showed that the outcomes of measurements for $d-1$ elements of a coset $%
A_{v}/C_{a}$, represented by $T_{rs}^{b}S_{ij}^{a}$ (with $b\neq
0$), are independent (see Theorem \ref{theorem LI}). Now we show
that measuring any
other operator, $T_{rs}^{b}S_{ij}^{a^{\prime }}$, from any other coset $%
A_{v}/C_{a^{\prime }}$, results in linearly dependent equations for
the
functions $w=$\textrm{tr}$(T_{rs}^{b}S_{ij}^{a}\rho _{k})$ and $w^{\prime }=$%
\textrm{tr}$(T_{rs}^{b}S_{ij}^{a^{\prime }}\rho _{k})$ as the
following:
\begin{widetext}
\begin{eqnarray*}
&&w = \mathrm{Tr}(T_{rs}^{b}S_{ij}^{a}\rho _{k})= \sum_{m}\chi
_{mm}~\mathrm{Tr} (E_{m}^{A}{}^{\dagger
}T_{rs}^{b}S_{ij}^{a}E_{m}^{A}\rho ) +\sum_{m<n}[\chi
_{mn}~\mathrm{Tr}(E_{n}^{A}{}^{\dagger
}T_{rs}^{b}S_{ij}^{a}E_{m}^{A}\rho )+\chi _{mn}^{* }~\mathrm{Tr}%
(E_{m}^{A}{}^{\dagger }T_{rs}^{b}S_{ij}^{a}E_{n}^{A}\rho )]\\
&&w^{\prime } = \mathrm{Tr}(T_{rs}^{b}S_{ij}^{a^{\prime }}\rho _{k}) =
\sum_{m}\chi _{mm}~\mathrm{Tr}(E_{m}^{A}{}^{\dagger
}T_{rs}^{b}S_{ij}^{a^{\prime }}E_{m}^{A}\rho )+ \sum_{m<n}[\chi
_{mn}~\mathrm{Tr}(E_{n}^{A}{}^{\dagger}T_{rs}^{b}S_{ij}^{a^{\prime }}E_{m}^{A}\rho )+\chi _{mn}^{*}~\mathrm{tr%
}(E_{m}^{A}{}^{\dagger }T_{rs}^{b}S_{ij}^{a^{\prime }}E_{n}^{A}\rho
)].
\end{eqnarray*}
Using the commutation relations
$T_{rs}^{b}S_{ij}^{a}E_{m}^{A}=\omega
^{br_{m}+ai_{m}}E_{m}^{A}T_{rs}^{b}S_{ij}^{a},$ we obtain:
\begin{eqnarray*}
&&w = \sum_{m}\omega ^{br_{m}+ai_{m}}\chi
_{mm}~\mathrm{Tr}(T_{rs}^{b}\rho )+\sum_{m<n}[\omega ^{br_{m}+ai_{m}}\chi _{mn}~\mathrm{Tr}%
(E_{n}^{A}{}^{\dagger }E_{m}^{A}T_{rs}^{b}\rho )+\omega
^{br_{n}+ai_{n}}\chi _{mn}^{* }~\mathrm{Tr}(E_{m}^{A}{}^{\dagger
}E_{n}^{A}T_{rs}^{b}\rho )]\\
&&w^{\prime } = \sum_{m}\omega ^{br_{m}+a^{\prime }i_{m}}\chi _{mm}~\mathrm{%
tr}(T_{rs}^{b}\rho )+ \sum_{m<n}[\omega ^{br_{m}+a^{\prime }i_{m}}\chi _{mn}~\mathrm{Tr}%
(E_{n}^{A}{}^{\dagger }E_{m}^{A}T_{rs}^{b}\rho )+\omega
^{br_{n}+a^{\prime }i_{n}}\chi _{mn}^{*
}~\mathrm{Tr}(E_{m}^{A}{}^{\dagger }E_{n}^{A}T_{rs}^{b}\rho )],
\end{eqnarray*}
where we also used the fact that both $S_{ij}^{a}$ and
$S_{ij}^{a^{\prime }}$ are stabilizer elements. Since all of the
operators $E_{m}^{A}$ belong to the same $W_{k}^{i}$, we have
$i_{m}=i_{n}=k$, and obtain:{}
\begin{eqnarray*}
&&w = \omega ^{ak}\left(\sum_{m}\omega ^{br_{m}}\chi _{mm}~\mathrm{Tr}%
(T_{rs}^{b}\rho )+ \sum_{m<n}[\omega ^{br_{m}}\chi
_{mn}~\mathrm{Tr}(E_{n}^{A}{}^{\dagger
}E_{m}^{A}T_{rs}^{b}\rho )+\omega ^{br_{n}}\chi _{mn}^{* }~\mathrm{Tr}%
(E_{m}^{A}{}^{\dagger }E_{n}^{A}T_{rs}^{b}\rho )]\right)\\
&&w^{\prime } = \omega ^{a^{\prime }k}\left(\sum_{m}\omega ^{br_{m}}\chi _{mm}~%
\mathrm{Tr}(T_{rs}^{b}\rho )+ \sum_{m<n}[\omega ^{br_{m}}\chi
_{mn}~\mathrm{Tr}(E_{n}^{A}{}^{\dagger
}E_{m}^{A}T_{rs}^{b}\rho )+\omega ^{br_{n}}\chi _{mn}^{* }~\mathrm{Tr}%
(E_{m}^{A}{}^{\dagger }E_{n}^{A}T_{rs}^{b}\rho )]\right).
\end{eqnarray*}
\end{widetext}
Thus, we have $w^{\prime }\ =\omega ^{(a^{\prime }-a)k}w$, and
consequently the measurements of operators from other cosets $%
A_{v}/C_{a^{\prime }}$ do not provide any new information about
$\chi _{mn}$ beyond the corresponding measurements from the coset
$A_{v}/C_{a}$.
\end{proof}\\

For another proof of the optimality, based on fundamental limitation
of transferring information between two parties given by the Holevo
bound see Ref.~\cite{MasoudThesis}. In principle, one can construct a set of
\emph{non-Abelian} normalizer measurements, from different $A_{v}$, where $%
v=1,2,\ldots,d+1$, to obtain information about the off-diagonal
elements $\chi _{mn}$. However, determining the eigenvalues of a set
of noncommuting operators cannot be done via a single measurement.
Moreover, as mentioned above, by measuring the stabilizer and $d-1$
Abelian normalizers, one can obtain $\log _{2}d^{2}$ bits of
classical information, which is the maximum allowed by the Holevo
bound \cite{HolevoBound}. Therefore, other strategies involving
non-Abelian, or a mixture of Abelian and non-Abelian normalizer
measurements, cannot improve the above scheme. It should be noted
that there are several possible alternative sets of Abelian
normalizers that are equivalent for this task. we address this issue
in the next lemma.
\begin{lemma}
The number of alternative sets of Abelian normalizer measurements
that can provide optimal information about quantum dynamics, in one
ensemble measurement, is $d^{2}$.
\end{lemma}
\begin{proof}
We have $d+1$ Abelian normalizers $A_{v}$ (see Lemma \ref{d+1 of
A}). However, there are $d$ of them that contain unitary operators
that act
nontrivially on both qudit systems $A$ and $B$, i.e., $%
T_{rs}^{b}=(E_{r}^{A}E_{s}^{B})^{b}$, where $E_{r}^{A}\neq I$, $%
E_{s}^{B}\neq I$. Moreover, in each $A_{v}$ we have $d$ cosets (see Lemma %
\ref{d+1 of A}) that can be used for optimal characterization of $\chi _{mn}$%
. Overall, we have $d^{2}$ possible sets of Abelian normalizers that
are equivalent for our purpose.
\end{proof}\\
In the next section, we develop the algorithm further to obtain
complete information about the off-diagonal elements of the
superoperator by repeating the above scheme for different input
states.

\section{Repeating the algorithm for other stabilizer states}

\label{rep-algorithm}

we have shown that by performing one ensemble measurement one can obtain $%
d^{2}$ linearly independent equations for $\chi _{mn}$. However, a
complete characterization of quantum dynamics requires obtaining
$d^{4}-d^{2}$ independent real parameters of the superoperator (or
$d^{4}$ for non-trace preserving maps). we next show how one can
obtain complete information by appropriately rotating the input
state and repeating the above algorithm for a complete set of
rotations.

\begin{lemma}
The number of independent eigenkets for the error operator basis
$\{E_{j}\}$, where $j=1,2,\ldots,d^{2}-1$, is $d+1$. These eigenkets
are mutually unbiased.
\end{lemma}
\begin{proof}
We have $d^{2}-1$ unitary operators ,$E_{i}$. We note that the operators $%
E_{i}^{a}$ for all values of $1\leq a\leq d-1$ commute and have a
common eigenket. Therefore, overall we have $(d^{2}-1)/(d-1)=d+1$
independent
eigenkets. Moreover, it has been shown \cite{som-mub} that if a set of $%
d^{2}-1 $ traceless and mutually orthogonal $d\times d$ unitary
matrices can be partitioned into $d+1$ subsets of equal size, such
that the $d-1$ unitary operators in each subset commute, then the
basis of eigenvectors defined by these subsets are mutually
unbiased.
\end{proof}

Let us construct a set of $d+1$ stabilizer operators
$E_{i}^{A}E_{j}^{B}$, such that the following conditions hold: (a)
$E_{i}^{A}$,$E_{j}^{B}\neq I$, (b) $(E_{i}^{A})^{a}\neq $
$E_{i^{\prime }}^{A}$ for $i\neq i^{\prime }$ and $1\leq a\leq d-1$.
Then, by preparing the eigenstates of these $d+1$ independent
stabilizer operators, one at a time, and measuring the eigenvalues
of $S_{ij}$ and its corresponding $d-1$ normalizer operators
$T_{rs}^{b}S_{ij}^{a}\in A_{v}/C_{a}$,$~$one can obtain $(d+1)d^{2}$
\textit{linearly independent} equations to characterize the
superoperator's off-diagonal elements.\ The linear independence of
these
equations can be understood by noting that the eigenstates of all operators $%
E_{i}^{A}$ of these $d+1$ stabilizer operator $S_{ij}$ are mutually
unbiased (i.e., the measurements in these mutual unbiased bases are
maximally
non-commuting). For example the bases $\{\left\vert 0\right\rangle $,$%
\left\vert 1\right\rangle \}$, $\{(\left\vert +\right\rangle _{X}$,$%
\left\vert -\right\rangle _{X}\}$ and $\{\left\vert +\right\rangle _{Y}$,$%
\left\vert -\right\rangle _{Y}\}$ (the eigenstates of the Pauli
operators $Z$, $X$, and $Y$) are \emph{mutually unbiased}, i.e., the
inner products of each pair of elements in these bases have the same
magnitude. Then measurements in these bases are maximally
non-commuting \cite{wootters-mub}.

To obtain complete information about the quantum dynamical
coherence, we
again prepare the eigenkets of the above $d+1$ stabilizer operators $%
E_{i}^{A}E_{j}^{B}$, but after the stabilizer measurement we
calculate the expectation values of the operators $T_{r^{\prime
}s^{\prime }}^{b}S_{ij}^{a} $ belonging to other Abelian subgroups
$A_{v^{\prime }}/C_{a}$ of the normalizer, i.e., $A_{v^{\prime
}}\neq A_{v}$ . According to Lemma \ref{MUA} the bases of different
Abelian subgroups of the
normalizer are mutually unbiased, therefore, the expectation values of $%
T_{r^{\prime }s^{\prime }}^{b}S_{ij}^{a}$ and $T_{rs}^{b}S_{ij}^{a}$
from different Abelian subgroups $A_{v^{\prime }}$ and $A_{v}$ are
independent. In order to make the stabilizer measurements also
independent we choose a different superposition of logical basis in
the preparation of $d+1$ possible stabilizer state in each run.
Therefore in each of these measurements we can obtain at most
$d^{2}$ linearly independent equations. By repeating these
measurements for $d-1$ different $A_{v}$ over all $d+1$ possible
input stabilizer state, we obtain $(d+1)(d-1)d^{2}=d^{4}-d^{2}$
linearly independent equations, which suffice to fully characterize
all independent parameters of the superoperator's off-diagonal
elements. In the next section, we address the general properties of
these $d+1$ stabilizer states.

\section{General constraints on the stabilizer states}

\label{constraints-prep}

The restrictions on the stabilizer states $\rho $ can be expressed
as follows:

\begin{condition}
The state $\rho =\left\vert \phi _{ij}\right\rangle \left\langle
\phi _{ij}\right\vert $ is a non-separable pure state in the Hilbert
space of the two-qudit system $\mathcal{H}$. I.e., $\left\vert \psi
_{ij}\right\rangle _{AB}\neq \left\vert \phi \right\rangle
_{A}\otimes \left\vert \varphi \right\rangle _{B}$.
\end{condition}

\begin{condition}
The state $\left\vert \phi _{ij}\right\rangle $ is a stabilizer
state with a sole stabilizer generator $S_{ij}=E_{i}^{A}E_{j}^{B}$.
I.e., it satisfies $\ S_{ij}^{a}\left\vert \phi _{ij}\right\rangle
=\omega ^{ak}\left\vert \phi _{ij}\right\rangle $, where $k\in
\{0,1,\ldots,d-1\}$ denotes a fixed eigenvalue of $S_{ij}$, and
$a=1,\ldots,d-1$ labels $d-1$ nontrivial members of the stabilizer
group.
\end{condition}
The second condition specifies the stabilizer subspace, $V_{S}$,
that the state $\rho $ \ lives in, which is the subspace fixed by
all the elements of the stabilizer group with a fixed eigenvalues
$k$. More specifically, an
arbitrary state in the entire Hilbert space $\mathcal{H}$ can be written as $%
\left\vert \phi \right\rangle =\sum\limits_{u,u^{\prime
}=0}^{d-1}\alpha _{uu^{\prime }}\left\vert u\right\rangle
_{A}\left\vert u^{\prime
}\right\rangle _{B}$ where $\{\left\vert u\right\rangle \}$ and $%
\{\left\vert u^{\prime }\right\rangle \}$ are bases for the Hilbert
spaces of the qudits $A$ and $B$, such that $X^{q}\left\vert
u\right\rangle =\left\vert u+q\right\rangle $ and $Z^{p}\left\vert
u\right\rangle =\omega ^{pu}\left\vert u\right\rangle $. However, we
can expand $\left\vert \phi \right\rangle $ in another basis as
$\left\vert \phi \right\rangle =\sum\limits_{v,v^{\prime
}=0}^{d-1}\beta _{vv^{\prime }}\left\vert
v\right\rangle _{A}\left\vert v^{\prime }\right\rangle _{B}$, such that $%
X^{q}\left\vert v\right\rangle =\omega ^{qv}\left\vert v\right\rangle $ and $%
Z^{p}\left\vert v\right\rangle =\left\vert v+p\right\rangle $. Let
us consider a stabilizer state fixed under the action of a unitary operator $%
E_{i}^{A}E_{j}^{B}=(X^{A})^{q}(X^{B})^{q^{\prime
}}(Z^{A})^{p}(Z^{B})^{p^{\prime }}$ with eigenvalue $\omega ^{k}$.
Regardless of the basis chosen to expand $\left\vert \phi
_{ij}\right\rangle $, we should always have $S_{ij}\left\vert \phi
_{ij}\right\rangle =\omega ^{k}\left\vert \phi _{ij}\right\rangle $.
Consequently, we have the constraints $pu\oplus p^{\prime }u^{\prime
}=k$,$\ $for the stabilizer subspace $V_{S}$ spanned by the
$\{\left\vert u\right\rangle \otimes \left\vert u^{\prime
}\right\rangle \}$ basis, and $q(v\oplus p)\oplus
q^{\prime }(v^{\prime }\oplus p^{\prime })=k$, if $V_{S}$ is spanned by $%
\{\left\vert v\right\rangle \otimes \left\vert v^{\prime
}\right\rangle \}$ basis, where $\oplus $ is addition
$\mathrm{mod}(d)$. From these relations,
and also using the fact that the bases $\{\left\vert v\right\rangle \}$ and $%
\{\left\vert u\right\rangle \}$ are related by a unitary
transformation, one
can find the general properties of $V_{S}$ for a given stabilizer generator $%
E_{i}^{A}E_{j}^{B}$ and a given $k$.

We have already shown that the stabilizer states $\rho $ should also
satisfy the set of conditions \textrm{Tr}$[E_{n}^{A}{}^{\dagger
}E_{m}^{A}\rho ]\neq 0$ and $\mathrm{Tr}(E_{n}^{A}{}^{\dagger
}E_{m}^{A}T_{rs}^{b}\rho )\neq 0$ for all operators $E_{m}^{A}$
belonging to the same $W_{k}^{i}$,
where $T_{rs}^{b}$ ($0<b\leq d-1$) are the members of a particular coset $%
A_{v}/C_{a}$ of an Abelian subgroup, $A_{v}$, of the normalizer
$N(S)$. These relations can be expressed more compactly as:
\begin{condition}
For stabilizer state $\rho =\left\vert \phi _{ij}\right\rangle
\left\langle \phi _{ij}\right\vert \equiv\left\vert \phi
_{c}\right\rangle \left\langle \phi _{c}\right\vert $ and for all
$E_{m}^{A}\in W_{k}^{i}$ we have:
\begin{equation}
\left\langle \phi _{c}\right\vert E_{n}^{A}{}^{\dagger
}E_{m}^{A}T_{rs}^{b}\left\vert \phi _{c}\right\rangle \neq 0,
\end{equation}
where here $0\leq b\leq d-1$.
\end{condition}
Before developing the implications of the above formula for the
stabilizer states we give the following definition and lemma.
\begin{definition}
Let $\{\left\vert l\right\rangle _{L}\}$ be the logical basis of the
code space that is fixed by the stabilizer generator
$E_{i}^{A}E_{j}^{B}$. The stabilizer state in that basis can be
written as $\left\vert \phi _{c}\right\rangle
=\sum\limits_{l=0}^{d-1}\alpha _{l}\left\vert l\right\rangle _{L}$,
and all the normalizer operators, $T_{rs}$, can be
generated from tensor products of logical operations $\overline{X}$ and $%
\overline{Z}$ defined as $\overline{Z}\left\vert l\right\rangle
_{L}=\omega ^{l}\left\vert l\right\rangle _{L}$ and
$\overline{X}\left\vert l\right\rangle _{L}=\left\vert
l+1\right\rangle $. For example: $\left\vert
l\right\rangle _{L}=\left\vert k\right\rangle \left\vert k\right\rangle $, $%
\overline{Z}=Z\otimes I$ and $\overline{X}=X\otimes X$, where
$X\left\vert k\right\rangle =\left\vert k+1\right\rangle $ and
$Z\left\vert k\right\rangle =\omega ^{k}\left\vert k\right\rangle $.
\end{definition}
\begin{lemma}
For a stabilizer generator $E_{i}^{A}E_{j}^{B}\ $and all unitary operators $%
E_{m}^{A}\in W_{k}^{i}$, we always have $E_{n}^{A}{}^{\dagger
}E_{m}^{A}=\omega ^{c}\overline{Z}^{a}$, where $\overline{Z}$ is the
logical $Z$ operation acting on the code space and $a$ and $c$ are
integers.
\end{lemma}
\begin{proof}
Let us consider $E_{i}^{A}=X^{q_{i}}Z^{p_{i}}$, and two generic operators $%
E_{n}^{A}{}$ and $E_{m}^{A}$ that belong to $W_{k}^{i}$: $%
E_{m}^{A}=X^{q_{m}}Z^{p_{m}}$ and $E_{n}^{A}=X^{q_{n}}Z^{p_{n}}$.
From the definition of $W_{k}^{i}$ (see Definition \ref{definition
Wk}) we have $\
p_{i}q_{m}-q_{i}p_{m}=p_{i}q_{n}-q_{i}p_{n}=k~(\mathrm{mod}d)=k+td$.
We can
solve these two equations to get: $%
q_{m}-q_{n}=q_{i}(p_{m}q_{n}-q_{m}p_{n})/(k+td)$ and $%
p_{m}-p_{n}=p_{i}(p_{m}q_{n}-q_{m}p_{n})/(k+td)$. We also define $%
p_{m}q_{n}-q_{m}p_{n}=k^{\prime }+t^{\prime }d$. Therefore, we obtain $%
q_{m}-q_{n}=q_{i}a$ and $p_{m}-p_{n}=p_{i}a$, where we have
introduced
\begin{equation}
a=(k^{\prime }+t^{\prime }d)/(k+td).  \label{a of k and k'}
\end{equation}%
Moreover, we have $E_{n}^{A\dagger
}=X^{(t''d-q_{n})}Z^{(t''d-p_{n})}$ for some other integer $t''$.
Then we get:

\begin{eqnarray*}
E_{n}^{A\dagger }E_{m}^{A} &=& \omega ^{c}X^{(t'' d + q_{m} - q_{n}
)} Z^{(t''d+p_{m}-p_{n})} \\ &=& \omega ^{c} X^{(q_{m}-q_{n})}
Z^{(p_{m}-q_{n})} \\ &=& \omega ^{c}(X^{q_{i}}Z^{p_{i}})^{a},
\end{eqnarray*}\\
where $c = (t'' d - p_{n} ) ( t'' d + q_{m} - q_{n} )$. However, $%
X^{q_{i}}Z^{p_{i}}$ $\otimes I$ acts as logical $\overline{Z}$ on
the code subspace, which is the eigenstate of $E_{i}^{A}E_{j}^{B}$.
Thus, we obtain $E_{n}^{A}{}^{\dagger }E_{m}^{A}=\omega
^{c}\overline{Z}^{a}$.
\end{proof}\\
Based on the above lemma, for the case of $b=0$ we obtain
\begin{equation*}
\left\langle \phi _{c}\right\vert E_{n}^{A}{}^{\dagger
}E_{m}^{A}\left\vert \phi _{c}\right\rangle =\omega ^{c}\left\langle
\phi _{c}\right\vert \overline{Z}^{a}\left\vert \phi
_{c}\right\rangle =\omega ^{c}\sum_{l=0}^{d-1}\omega ^{al}\left\vert
\alpha _{l}\right\vert ^{2}.
\end{equation*}
Therefore, our constraint in this case becomes
$\sum_{k=0}^{d-1}\omega ^{al}\left\vert \alpha _{l}\right\vert
^{2}\neq 0,$ which is not satisfied
if the stabilizer state is maximally entangled. For $b\neq 0$, we note that $%
T_{rs}^{b}$ are in fact the normalizers. By considering the
general form of the normalizer elements as $T_{rs}^{b}=$ $(\overline{X}^{q}%
\overline{Z}^{p})^{b}$, where $q$, $p\in \{0,1,\ldots,d-1\}$, we
obtain:
\begin{eqnarray*}
\left\langle \phi _{c}\right\vert E_{n}^{A}{}^{\dagger
}E_{m}^{A}T_{rs}^{b}\left\vert \phi _{c}\right\rangle &=& \omega
^{c}\left\langle \phi _{c}\right\vert \overline{Z}^{a}(\overline{X}^{q}%
\overline{Z}^{p})^{b}\left\vert \phi _{c}\right\rangle \\ &=& \omega
^{c}\sum_{k=0}^{d-1}\omega ^{a(l+bq)}\omega ^{bpl}\alpha _{l}^{\ast
}\alpha
_{l+bq} \\
&=& \omega ^{(c+abq)}\sum_{l=0}^{d-1}\omega ^{(a+bp)l}\alpha
_{l}^{\ast }\alpha _{l+bq}.
\end{eqnarray*}
Overall, the constraints on the stabilizer state, due to condition
(iii), can be summarized as:
\begin{equation}
\sum_{l=0}^{d-1}\omega ^{(a+bp)l}\alpha _{l}^{\ast }\alpha
_{l+bq}\neq 0 \label{SCodesConditions}
\end{equation}%
This inequality should hold for all $b\in \{0,1,\ldots,d-1\}$, and
all $a$
defined by Eq.~(\ref{a of k and k'}), however, for a particular coset $%
A_{v}/C_{a}$ the values of $q$ and $p$ are fixed. One important
property of the stabilizer code, implied by the above formula with
$b=0$, is that it should always be a \textit{nonmaximally entangled
state. }In the next section, by utilizing the quantum Hamming bound,
we show that the minimum number of physical qudits, $n$, needed for
encoding the required stabilizer state is in fact \textit{two}.

\section{Minimum number of required physical qudits}

\label{minimum-qudits}

In order to characterize off-diagonal elements of a superoperator we
have to use degenerate stabilizer codes, in order to preserve the
coherence between operator basis elements. Degenerate stabilizer
codes do not have a classical analog \cite{nielsen-book}. Due to
this fact, the classical techniques used to prove bounds for
non-degenerate error-correcting codes cannot be applied to
degenerate codes. In general, it is yet unknown if there are
degenerate codes that exceed the quantum Hamming bound
\cite{nielsen-book}. However, due to the simplicity of the
stabilizer codes used in the DCQD algorithm and their symmetry, it
is possible to generalize the quantum Hamming bound for them. Let us
consider a stabilizer code that is used for encoding $k$ logical
qudits into $n$ physical qudits such that we can correct any subset
of $t$ or fewer errors on any $n_{e}\leqslant n$ of the physical qudits. Suppose that $%
0\leqslant j\leqslant t$ errors occur. Therefore, there are $\left(
\begin{smallmatrix}
n_{e} \\
j%
\end{smallmatrix}%
\right) $ possible locations, and in each location there are
$(d^{2}-1)$ different operator basis elements that can act as
errors. The total possible number of errors is $\sum_{j=0}^{t}\left(
\begin{smallmatrix}
n_{e} \\
j%
\end{smallmatrix}%
\right) (d^{2}-1)^{j}.$ If the stabilizer code is non-degenerate
each of these errors should correspond to an orthogonal
$d^{k}$-dimensional subspace; but if the code is uniformly $g$-fold
degenerate (i.e., with respect to all possible errors), then each
set of $g$ errors can be fit into an orthogonal $d^{k}$-dimensional
subspace. All these subspaces must be fit into the entire
$d^{n}$-dimensional Hilbert space. This leads to the following
inequality:
\begin{equation}
\sum_{j=0}^{t}\left(
\begin{array}{c}
n_{e} \\
j%
\end{array}%
\right) \frac{(d^{2}-1)^{j}d^{k}}{g}\leq d^{n}.  \label{Hamming}
\end{equation}%
We are always interested in finding the errors on one physical
qudit. Therefore, we have $n_{e}=1,$ $j\in \{0,1\}$ and $\left(
\begin{smallmatrix}{c}
n_{e} \\
j%
\end{smallmatrix}%
\right) =1$, and Eq.~(\ref{Hamming}) becomes $\sum_{j=0}^{1}\frac{%
(d^{2}-1)^{j}d^{k}}{g}\leq d^{n}.$ In order to characterize diagonal
elements, we
use a nondegenerate stabilizer code with $n=2,~k=0$ and $g=1,$ and we have $%
\sum_{j=0}^{1}(d^{2}-1)^{j}=d^{2}$. For off-diagonal elements, we
use a
degenerate stabilizer code with $n=2,~k=1$ and $g=d,$ and we have $%
\sum_{j=0}^{1}\frac{(d^{2}-1)^{j}d}{d}=d^{2}.$ Therefore, in the
both cases the upper-bound of the quantum Hamming bound is satisfied
by our codes. Note that if instead we use $n=k,$ i.e., if we encode
$n$ logical qudits into $n$ separable physical qubits, we get
$\sum_{j=0}^{1}\frac{(d^{2}-1)^{j}}{g}\leq 1.$ This can only be
satisfied if $g=d^{2},\,\ $in which case we cannot obtain any
information about the errors. The above argument justifies Condition
(i) of the stabilizer state being nonseparable. Specifically, it
explains why alternative encodings such as $n=k=2$ and $n=k=1$ are
excluded from our discussions. However, if we encode zero logical
qubits into one physical qubit, i.e., $n=1,~k=0,$ then, by using a
$d$-fold degenerate code, we can obtain
$\sum_{j=0}^{1}\frac{(d^{2}-1)^{j}}{d}=d$ which satisfies the
quantum Hamming bound and could be useful for characterizing
off-diagonal elements. For this to be true, the code $\left\vert
\phi _{c}\right\rangle $ should also satisfy the set of conditions
$\left\langle \phi _{c}\right\vert E_{n}^{A}{}^{\dagger
}E_{m}^{A}\left\vert \phi _{c}\right\rangle \neq 0$ and
$\left\langle \phi _{c}\right\vert E_{n}^{A}{}^{\dagger
}E_{m}^{A}T_{rs}^{b}\left\vert \phi _{c}\right\rangle \neq 0.$ Due to the $d$%
-fold degeneracy of the code, the condition $\left\langle \phi
_{c}\right\vert E_{n}^{A}{}^{\dagger }E_{m}^{A}\left\vert \phi
_{c}\right\rangle \neq 0$ is automatically satisfied. However, the
condition $\left\langle \phi _{c}\right\vert E_{n}^{A}{}^{\dagger
}E_{m}^{A}T_{rs}^{b}\left\vert \phi _{c}\right\rangle \neq 0$ can
never be satisfied, since the code space is one-dimensional, i.e.,
$d^{k}=1,$ and the normalizer operators cannot be defined. I.e.,
there does not exist any nontrivial unitary operator $T_{rs}^{b}$
that can perform logical operations on the one-dimensional code
space.

we have demonstrated how we can characterize quantum dynamics using
the most general form of the relevant stabilizer states and
generators. In the next section, we choose a standard form of
stabilizers, in order to simplify the algorithm and to derive a
standard form of the normalizer.

\section{Standard form of stabilizer and normalizer operators}

\label{standar-form-S}

Let us choose the set $\{\left\vert 0\right\rangle ,\left\vert
1\right\rangle ,...,\left\vert k-1\right\rangle \}$ as a standard
basis, such that $Z\left\vert k\right\rangle =\omega ^{k}\left\vert
k\right\rangle $ and$~X\left\vert k\right\rangle =\left\vert
k+1\right\rangle $. In order to characterize the quantum dynamical
population, we choose the standard stabilizer generators to be
$(X^{A}X^{B})^{q}$ and$\ [Z^{A}(Z^{B})^{d-1}]^{p} $. Therefore, the
maximally entangled input states
can be written as $\left\vert \varphi _{c}\right\rangle =\frac{1}{\sqrt{d}}%
\sum\limits_{k=0}^{d-1}\left\vert k\right\rangle _{A}\left\vert
k\right\rangle _{B}$. In order to characterize the quantum dynamical
coherence we choose the sole stabilizer operator as
$[E_{i}^{A}(E_{i}^{B})^{d-1}]^{a},$ which has an eigenket of the
form $\left\vert \varphi _{c}\right\rangle
=\sum\limits_{i=0}^{d-1}\alpha _{i}\left\vert i\right\rangle
_{A}\left\vert i\right\rangle _{B},$ where $E_{i}\left\vert
i\right\rangle =\omega ^{i}\left\vert i\right\rangle $ and
$\left\vert i\right\rangle $ represents one of $d+1$ mutually
unbiased basis states in the Hilbert space of one
qudit. The normalizer\ elements can be written as $T_{qp}^{b}=$ $(\overline{X%
}^{q}\overline{Z}^{p})^{b}\in A_{v_{0}}/C_{a_{0}},$ for all $0<b\leq
d-1,$
where $\overline{X}=\widetilde{E_{i}}\otimes \widetilde{E_{i}}$ $,$ $%
\overline{Z}=E_{i}\otimes I,$ $\widetilde{E_{i}}\left\vert
i\right\rangle =\left\vert i+1\right\rangle $ and $E_{i}\left\vert
i\right\rangle =\omega ^{i}\left\vert i\right\rangle ;$ and
$A_{v_{0}}/C_{a_{0}}$ represents a fixed coset of a particular
Abelian subgroup, $A_{v_{0}},$ of the normalizer
$N(S)$. For example, for a stabilizer generator of the form $%
[E_{i}^{A}(E_{i}^{B})^{d-1}]^{a}=[Z^{A}(Z^{B})^{d-1}]^{p}$ \ we
prepare its eigenket $\left\vert \varphi _{c}\right\rangle
=\sum\limits_{k=0}^{d-1}\alpha _{k}\left\vert k\right\rangle
_{A}\left\vert
k\right\rangle _{B},$ and the normalizers become $T_{qp}^{b}=$ $(\overline{X}%
^{q}\overline{Z}^{p})^{b},$ where $\overline{X}=X\otimes X$ and $\overline{Z}%
=Z\otimes I.$ Using this notations for stabilizer and the normalizer
operators, we provide an overall outline for the DCQD algorithm in
the next section.

\section{Algorithm: Direct characterization of quantum dynamics}

\label{algorithm-summary}

\begin{center}
\begin{figure*}[tp]
\begin{center}
\includegraphics[width=9.5cm,height=4cm]{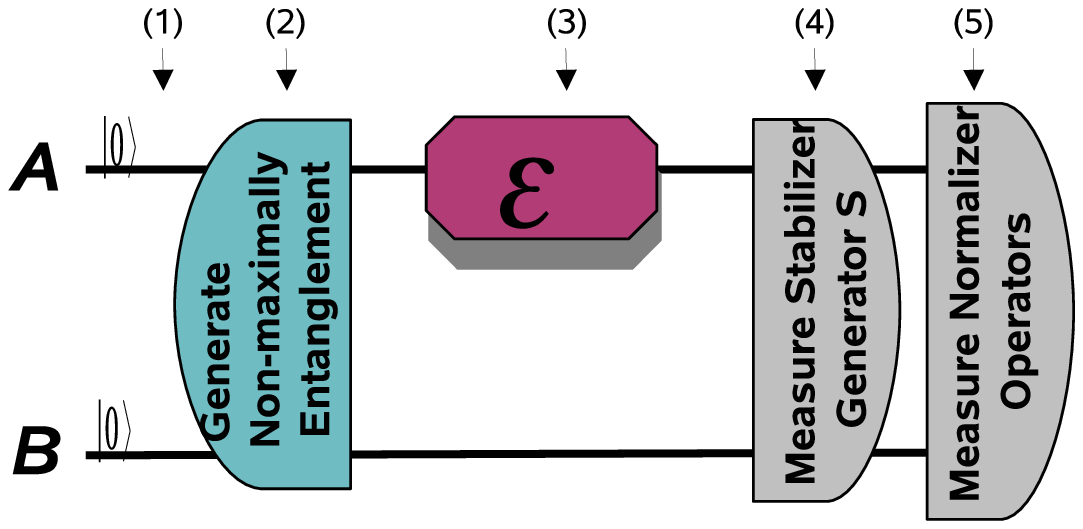}
\end{center}
\caption[\textbf{Procedure (b)}: Measuring the quantum dynamical
coherence (off-diagonal elements $\chi
_{mn}$).]{\small{\textbf{Procedure (b)}: Measuring the quantum
dynamical coherence (off-diagonal elements $\chi _{mn}$). (1)
Prepare $\left\vert \varphi _{0}\right\rangle =\left\vert
0\right\rangle _{A}\otimes \left\vert 0\right\rangle _{B}$, a
pure initial state. (2) Transform it to $\left\vert \varphi %
_{C}\right\rangle =\sum\limits_{i=0}^{d-1}\alpha _{i}\left\vert
i\right\rangle _{A}\left\vert i\right\rangle _{B}$, a
\textit{nonmaximally} entangled state of the two qudits. This state
has a sole stabilizer operator of the form
$S=E_{i}^{A}(E_{i}^{B})^{d-1}$ . (3) Apply the unknown quantum
dynamical map to the qudit $A$, $\mathcal{E}(\rho )$, where $%
\rho =\left\vert \phi _{C}\right\rangle \left\langle \phi %
_{C}\right\vert $. (4) Perform a projective measurement $P_{k}$, and
calculate the probability of the outcome $k$:\ \textrm{tr}$[P_{k}\mathcal{E}(%
\rho )]$. (5) Measure the expectation values of all normalizer
operators $T_{rs}$ that simultaneously commute with the stabilizer
generator $S$. There are only $d-1$ such operators $T_{rs}$ that are
independent of each other, within a multiplication by a stabilizer
generator; and they belong to an Abelian subgroup of the normalizer
group. (6) Repeat the steps (1)-(5) $d+1$ times, by preparing the
eigenkets of other stabilizer operator $E_{i}^{A}(E_{i}^{B})^{d-1}$
for all $i\in \{1,2,\ldots, d+1\}$, such that states $\left\vert
i\right\rangle _{A}$ in step (2) belong to a mutually unbiased basis
\cite{wootters-mub}. (7) Repeat the step (6) up to $d-1$ times, each
time choosing normalizer elements $T_{rs}$ from a different Abelian
subgroup of the normalizer, such that these measurements become
maximally non-commuting, i.e., their eigenstates form a set of
mutually unbiased bases. \emph{The number of ensemble measurements
for Procedure (b)} is $(d+1)(d-1)$.}}
\label{FigDCQD-general-algorithm2} 
\end{figure*}
\end{center}

The DCQD algorithm for the case of a qudit system is summarized as
follows
(see also Figs.~\ref{FigDCQD-general-algorithm1} and \ref%
{FigDCQD-general-algorithm2}.):

\begin{widetext}
\noindent \emph{\textbf{Inputs:}} (1) An ensemble of two-qudit
systems, $A$ and $B$, prepared in the state $\left\vert
0\right\rangle _{A}\otimes \left\vert 0\right\rangle _{B}$. (2) An
arbitrary unknown CP quantum
dynamical map $\mathcal{E}$, whose action can be expressed by $\mathcal{E}%
(\rho )=\sum_{m,n=0}^{d^{2}-1}\chi _{mn}~E_{m}^{A}\rho
E_{n}^{A\dagger }$, where $\rho $ denotes the state of the primary
system and the ancilla.\\
\noindent \emph{\textbf{Output:}} $\mathcal{E}$, given by a set of
measurement outcomes in the procedures (a) and (b)\ below:\\
\noindent\emph{\textbf{Procedure(a):}} Characterization of quantum
dynamical population (diagonal elements $\chi _{mm}$ of $\chi $),
see Fig.~\ref{FigDCQD-general-algorithm1}.

1. Prepare $\left\vert \varphi _{0}\right\rangle =\left\vert
0\right\rangle _{A}\otimes \left\vert 0\right\rangle _{B}$, a pure
initial state.

2. Transform it to $\left\vert \varphi _{c}\right\rangle =\frac{1}{\sqrt{d}}%
\sum\limits_{k=0}^{d-1}\left\vert k\right\rangle _{A}\left\vert
k\right\rangle _{B}$, a maximally entangled state of the two qudits.
This
state has the stabilizer operators $E_{i}^{A}E_{j}^{B}=(X^{A}X^{B})^{q}$ and$%
\ E_{i^{\prime }}^{A}E_{j^{\prime }}^{B}=[Z^{A}(Z^{B})^{d-1}]^{p}$ for $0<p$,%
$q\leq d-1$.

3. Apply the unknown quantum dynamical map to the qudit $A$:
$\mathcal{E}(\rho
)=\sum_{m,n=0}^{d^{2}-1}\chi _{mn}~E_{m}^{A}\rho E_{n}^{A\dagger }$, where $%
\rho =\left\vert \phi _{c}\right\rangle \left\langle \phi
_{c}\right\vert $.

4. Perform a projective measurement $P_{k}P_{k^{\prime }}:
\mathcal{E}(\rho )\mapsto P_{k}P_{k^{\prime }}\mathcal{E}(\rho
)P_{k}P_{k^{\prime }}$, where $
P_{k}=\frac{1}{d}\sum_{l=0}^{d-1}\omega
^{-lk}(E_{i}^{A}E_{j}^{B})^{l}$ , and $P_{k^{\prime
}}=\frac{1}{d}\sum_{l^{\prime }=0}^{d-1}\omega ^{-l^{\prime
}k^{\prime }}(E_{i^{\prime }}^{A}E_{j^{\prime
}}^{B})^{l^{\prime }},$ and calculate the joint probability distributions of the outcomes $k$ and $%
k^{\prime }$:
\begin{eqnarray*}
\mathrm{Tr} [P_{k}P_{k^{\prime }}\mathcal{E}(\rho )]=\chi _{mm}.
\end{eqnarray*}

\emph{Number of ensemble measurements for Procedure (a)}: $1$.

\begin{center}
\begin{figure*}[tp]
\begin{center}
\includegraphics[width=8cm,height=4.1cm]{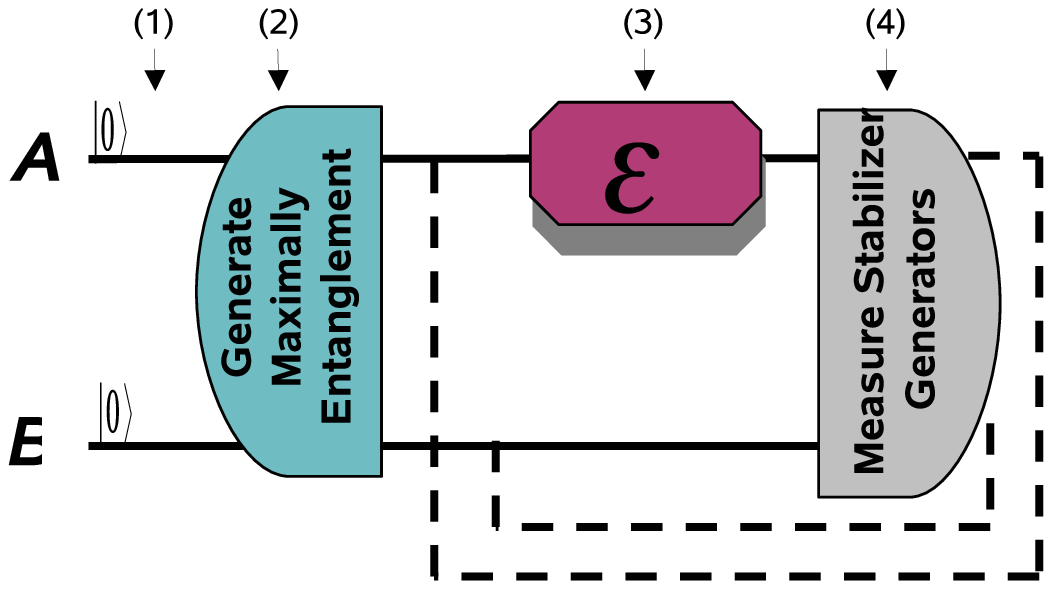}
\end{center}
\caption[\textbf{Procedure (a)}:
Measuring the quantum dynamical population (diagonal elements $\chi %
_{mm}$)]{\small{\textbf{Procedure (a)}:
Measuring the quantum dynamical population (diagonal elements $\chi %
_{mm}$). The arrows indicate direction of time. (1) Prepare
$\left\vert \varphi _{0}\right\rangle =\left\vert 0\right\rangle
_{A}\otimes \left\vert 0\right\rangle _{B}$, a pure initial state.
(2) Transform it to $\left\vert
\varphi _{C}\right\rangle =\frac{1}{\sqrt{d}}%
\sum\limits_{k=0}^{d-1}\left\vert k\right\rangle _{A}\left\vert
k\right\rangle _{B}$, a maximally entangled state of the two qudits.
This state has the stabilizer operators $S=X^{A}X^{B}$ and$\
S^{\prime
}=Z^{A}(Z^{B})^{d-1}$. (3) Apply the unknown quantum dynamical map to the qudit $%
A$, $\mathcal{E}(\rho )$, where $\rho =\left\vert \phi
_{C}\right\rangle \left\langle \phi _{C}\right\vert $. (4) Perform a
projective measurement $P_{k}P_{k^{\prime }},$ for $k,k^{\prime
}=0,\ldots,d-1,$ corresponding to eigenvalues of the stabilizer operators $S$ and$%
\ S^{\prime }$. Then calculate the joint probability distributions
of the
outcomes $k$ and $k^{\prime }$: $\mathrm{Tr}[P_{k}P_{k^{\prime }}\mathcal{E}(%
\rho )]=\chi _{mm}$. The elements $\chi _{mm}$ represent the
population of error operators that anticommute with the stabilizer
generators $S$ and $S^{\prime }$ with eigenvalues $\omega ^{k}$ and $%
\omega ^{k^{\prime }}$, respectively. The number of ensemble
measurements for procedure (a) is \emph{one}.}}
\label{FigDCQD-general-algorithm1} 
\end{figure*}
\end{center}

\noindent \emph{\textbf{Procedure (b):}} Characterization of quantum
dynamical coherence (off-diagonal elements $\chi _{mn}$ of $\chi $),
see Fig. \ref{FigDCQD-general-algorithm2}.

1. Prepare $\left\vert \varphi _{0}\right\rangle =\left\vert
0\right\rangle _{A}\otimes \left\vert 0\right\rangle _{B}$, a pure
initial state.

2. Transform it to $\left\vert \varphi _{c}\right\rangle
=\sum\limits_{i=0}^{d-1}\alpha _{i}\left\vert i\right\rangle
_{A}\left\vert i\right\rangle _{B}$, a non-maximally entangled state
of the two qudits. This state has stabilizer operators
$[E_{i}^{A}(E_{i}^{B})^{d-1}]^{a}$.

3. Apply the unknown quantum dynamical map to the qudit $A$:
$\mathcal{E}(\rho
)=\sum_{m,n=0}^{d^{2}-1}\chi _{mn}~E_{m}^{A}\rho E_{n}^{A\dagger }$, where $%
\rho =\left\vert \phi _{c}\right\rangle \left\langle \phi
_{c}\right\vert $.

4. Perform a projective measurement
\begin{eqnarray*}
P_{k}: \mathcal{E}(\rho )\mapsto \rho _{k}=P_{k}\mathcal{E}(\rho
)P_{k}=\sum\limits_{m}\chi _{mm}~E_{m}^{A}\rho E_{m}^{A\dagger
}+\sum_{m<n}(\chi _{mn}~E_{m}^{A}\rho E_{n}^{A\dagger }+\chi
_{mn}^{* }~E_{n}^{A}\rho E_{m}^{A\dagger }),
\end{eqnarray*}
where $P_{k}=\frac{1}{d}\sum_{l=0}^{d-1}\omega
^{-lk}(E_{i}^{A}E_{j}^{B})^{l} $ and
$E_{m}^{A}=X^{q_{m}}Z^{p_{m}}\in W_{k}^{i}$, and calculate the
probability of outcome $k$:
\begin{equation}
\mathrm{Tr}[P_{k}\mathcal{E}(\rho )]\ =\sum_{m}\chi _{mm}+2\sum_{m<n}\mathrm{%
Re}[\chi _{mn}~\mathrm{Tr}(E_{n}^{A}{}^{\dagger }E_{m}^{A}\rho )].
\end{equation}

5. Measure the expectation values of the normalizer\ operators
$T_{qp}^{b}=$
$(\overline{X}^{q}\overline{Z}^{p})^{b}\in A_{v_{0}}/C_{a_{0}}$, for all $%
0<b\leq d-1$, where $\overline{X}=\widetilde{E_{i}}\otimes
\widetilde{E_{i}}$ , $\overline{Z}=E_{i}\otimes I$,$~E_{i}\left\vert
i\right\rangle =\omega ^{i}\left\vert i\right\rangle $,
$\widetilde{E_{i}}\left\vert i\right\rangle =\left\vert
i+1\right\rangle $, where $A_{v_{0}}/C_{a_{0}}$ represents a fixed
coset of a particular Abelian subgroup, $A_{v_{0}}$, of the
normalizer $N(S)$.
\begin{eqnarray*}
\mathrm{Tr}(T_{qp}^{b}\rho _{k}) &=& \sum_{m}\omega
^{pq_{m}-qp_{m}}\chi _{mm}~\mathrm{Tr}(T_{rs}^{b}\rho )+ \sum_{m<n}[\omega ^{pq_{m}-qp_{m}}\chi _{mn}~\mathrm{Tr}%
(E_{n}^{A}{}^{\dagger }E_{m}^{A}T_{rs}^{b}\rho )+\omega
^{pq_{n}-qp_{n}}\chi _{mn}^{* }~\mathrm{Tr}(E_{m}^{A}{}^{\dagger
}E_{n}^{A}T_{rs}^{b}\rho )].
\end{eqnarray*}

6. Repeat the steps (1)-(5) $d+1$ times, by preparing the eigenkets
of other stabilizer operator $[E_{i}^{A}(E_{i}^{B})^{d-1}]^{a}$ for
all $i\in \{1,2,...,d+1\}$, such that states $\left\vert
i\right\rangle _{A}\left\vert i\right\rangle _{B}$ in the step (2)
belong to a mutually unbiased basis.

7. Repeat the step (6) up to $d-1$ times, each time choosing
normalizer\ elements $T_{qp}^{b}$ from a different Abelian subgroup
$A_{v}/C_{a}$, such that these measurements become maximally
non-commuting.

\emph{Number of ensemble measurements for Procedure (b)}:
$(d+1)(d-1)$.

\emph{Overall number of ensemble measurements}: $d^{2}$.
\end{widetext}

Note that at the end of each measurement in Figs.~\ref%
{FigDCQD-general-algorithm1} and \ref{FigDCQD-general-algorithm2},
the output state -- a maximally entangled state, $\left\vert \varphi
_{E}\right\rangle =\sum\limits_{i=0}^{d-1}\left\vert i\right\rangle
_{A}\left\vert i\right\rangle _{B}$ -- is the common eigenket of the
stabilizer generator and its commuting normalizer operators. For the
procedure (a), this state can be directly used for other
measurements. This is indicated by the dashed lines in
Fig.~\ref{FigDCQD-general-algorithm1}. For the procedure (b), the
state $\left\vert \varphi _{E}\right\rangle $ can be
unitarily transformed to another member of the same input stabilizer code, $%
\left\vert \varphi _{C}\right\rangle =\sum\limits_{i=0}^{d-1}\alpha
_{i}\left\vert i\right\rangle _{A}\left\vert i\right\rangle _{B}$,
before another measurement. Therefore, all the required ensemble
measurements, for measuring the expectation values of the stabilizer
and normalizer operators, can always be performed in a temporal
sequence on the same pair of qudits.

In the previous sections, we have explicitly shown how the DCQD
algorithm can be developed for qudit systems when $d$ is prime. In
the appendix \ref{generalization-Algorithm}, we demonstrate that the
DCQD algorithm can be generalized to other $N$-dimensional quantum
systems with $N$ being a power of a prime.

\section{Summary}

\label{summary-qudit}

For convenience, we provide a summary of the DCQD algorithm. The DCQD algorithm for a qudit, with $d$ being a prime, was
developed by
utilizing the concept of an error operator basis. An arbitrary operator acting on a qudit can be
expanded
over an orthonormal and unitary operator basis $\{E_{0}$,$E_{1}$,$\ldots$,$%
E_{d^{2}-1}\}$, where $E_{0}=I$ and \textrm{tr}$(E_{i}^{\dagger
}E_{j})=d\delta _{ij}$. Any element $E_{i}$ can be generated from
tensor products of $X$ and $Z$, where $X\left\vert k\right\rangle
=\left\vert k+1\right\rangle $ and $Z\left\vert k\right\rangle
=\omega ^{k}\left\vert k\right\rangle $, such that the relation
$XZ=\omega ^{-1}ZX$ is satisfied
\cite{Gottessmanhighd}. Here $\omega $ is a $d$th root of unity and $X$ and $%
Z$ are the generalizations of Pauli operators to higher dimension.

\textit{Characterization of Dynamical Population}.-- A measurement
scheme for determining the quantum dynamical population, $\chi
_{mm}$, in a single experimental configuration. Let us prepare a
maximally entangled state of
the two qudits $\left\vert \varphi _{C}\right\rangle =\frac{1}{\sqrt{d}}%
\sum\limits_{k=0}^{d-1}\left\vert k\right\rangle _{A}\left\vert
k\right\rangle _{B}$. This state is stabilized under the action of
stabilizer operators $S=X^{A}X^{B}$ and$\ S^{\prime
,}=Z^{A}(Z^{B})^{d-1}$,
and it is referred to as a \emph{stabilizer state} \cite%
{nielsen-book,Gottessmanhighd}. After applying the quantum map to
the qudit $A$, $\mathcal{E}(\rho )$, where $\rho =\left\vert \phi
_{C}\right\rangle \left\langle \phi _{C}\right\vert $, we can
perform a projective measurement
$P_{k}P_{k^{\prime }}\mathcal{E}(\rho )P_{k}P_{k^{\prime }}$, where $P_{k}=%
\frac{1}{d}\sum_{l=0}^{d-1}\omega ^{-lk}S^{l}$, $P_{k^{\prime }}=\frac{1}{d}%
\sum_{l^{\prime }=0}^{d-1}\omega ^{-l^{\prime }k^{\prime }}S^{\prime
l^{\prime }}$, and $\omega =e^{i2\pi /d}$. Then, we calculate the
joint
probability distributions of the outcomes $k$ and $k^{\prime }$: $\mathrm{Tr}%
[P_{k}P_{k^{\prime }}\mathcal{E}(\rho )]=\chi _{mm}$, where the elements $%
\chi _{mm}$ represent the population of error operators that
anticommute with stabilizer generators $S$ and $S^{\prime }$ with
eigenvalues $\omega
^{k}$ and $\omega ^{k^{\prime }}$, respectively. Therefore, with a \emph{%
single} experimental configuration we can identify all diagonal
elements of superoperator.

\textit{Characterization of Dynamical Coherence}.-- For measuring
the quantum dynamical coherence, we create a \textit{nonmaximally}
entangled state of the two qudits $\left\vert \varphi
_{C}\right\rangle =\sum\limits_{i=0}^{d-1}\alpha _{i}\left\vert
i\right\rangle _{A}\left\vert
i\right\rangle _{B}$. This state has the sole stabilizer operator $%
S=E_{i}^{A}(E_{i}^{B})^{d-1}$ (for detailed restrictions on the
coefficients $\alpha _{i}$ see Sec.~\ref{constraints-prep}). After
applying the dynamical map
to the qudit $A$, $\mathcal{E}(\rho )$, we perform a projective measurement $%
\rho _{k}=P_{k}\mathcal{E}(\rho )P_{k}$, and calculate the
probability of the outcome $k$: $\mathrm{Tr}[P_{k}{\mathcal{E}}(\rho
)]=\sum_{m}\chi _{mm}+2\sum_{m<n}\mathrm{Re}[\chi
_{mn}~\mathrm{Tr}(E_{n}^{A}{}^{\dagger }E_{m}^{A}\rho )]$; where
$E_{m}^{A}$ are all the operators in the operator basis,
$\{E_{j}^{A}\}$, that anticommute with the operator $E_{i}^{A}$ with
the same eigenvalue $\omega ^{k}$. We also measure the expectation
values of all independent operators $T_{rs}=E_{r}^{A}E_{s}^{B}$ of
the Pauli group (where $E_{r}^{A}\neq I$; $E_{s}^{B}\neq I)$ that
simultaneously commute with the stabilizer generator $S$:
$\mathrm{Tr}(T_{rs}\rho _{k})$. There are only $d-1$ such operators
$T_{rs}$ that are independent of each other, within a multiplication
by a stabilizer generator; and they belong to an Abelian subgroup of
the normalizer group. The normalizer group is the group of unitary
operators that preserve the stabilizer group by conjugation, i.e.,
$TST^{\dagger }=S$. We repeat this procedure $d+1$ times, by
preparing the eigenkets of other stabilizer operator
$E_{i}^{A}(E_{i}^{B})^{d-1}$ for all $i\in \{1,2,\ldots,d+1\}$, such
that states $\left\vert i\right\rangle _{A}$ in input states belong
to a mutually unbiased basis \cite{wootters-mub}. Also, we can
change the measurement basis $d-1$ times, each time choosing
normalizer elements $T_{rs}$ from a different Abelian subgroup of
the normalizer, such that their eigenstates form a mutually unbiased
basis in the code space. Therefore, we can completely characterize
quantum dynamical coherence by $(d+1)(d-1)$ different measurements,
and the overall number number of experimental configuration for a
qudit becomes $d^{2}$. For $N$-dimensional quantum systems, with $N$
a power of a prime, the required measurements are simply the tensor
product of the corresponding measurements on individual qudits -- see
Appendix~\ref{generalization-Algorithm}. For quantum system whose
dimension is not a power of a prime, the task can be accomplished by
embedding the system in a larger Hilbert space whose dimension is a
prime.

\section{Outlook}
\label{outlook}

An important and promising advantage of DCQD is for use in \textit{%
partial} characterization of quantum dynamics, where in general, one
cannot afford or does not need to carry out a full characterization
of the quantum system under study, or when one has some \textit{a
priori} knowledge about the dynamics. Using indirect methods of QPT
in those situations is inefficient, because one has to apply the
whole machinery of the scheme to obtain the relevant information
about the system. On the other hand, the DCQD scheme has built-in
applicability to the task of partial characterization of quantum
dynamics. In general, one can substantially reduce the overall
number of measurements, when estimating the coherence elements of
the superoperator for only specific subsets of the operator basis
and/or subsystems of interest. This fact has been demonstrated in
Ref.~\cite{MasoudThesis} in a generic fashion, and several examples
of partial characterization have also been presented. Specifically,
it was shown that DCQD can be efficiently applied to (single- and
two-qubit) Hamiltonian identification tasks. Moreover, it is
demonstrated that the DCQD algorithm enables the simultaneous
determination of coarse-grained (semiclassical) physical quantities,
such as the longitudinal relaxation time $T_{1}$ and the transversal
relaxation (or dephasing) time $T_{2}$ for a single qubit undergoing
a general CP quantum map. The DCQD scheme can also be used for
performing generalized quantum dense coding tasks. Other
implications and applications of DCQD for partial QPT remain to be
investigated and explored.

An alternative representation of the DCQD scheme for
higher-dimensional quantum systems, based on generalized Bell-state
measurements will be presented in Ref.~\cite{MasoudAli07}. The
connection of Bell-state measurements to stabilizer and normalizer
measurements in DCQD for two-level systems, can be easily observed
from Table II of Ref.~\cite{mohseni-rezakhani-lidar07}. Our
presentation of the DCQD algorithm assumes ideal (i.e., error-free)
quantum state preparation, measurement, and ancilla channels.
However, these assumptions can all be relaxed in certain situations,
in particular when the imperfections are already known. A discussion
of these issues is beyond the scope of this work and will be the
subject of a future publication \cite{MasoudAli07}.

There are a number of other directions in which the results
presented here can be extended. One can combine the DCQD algorithm
with the method of maximum likelihood estimation \cite{Kosut04}, in
order to minimize the statistical errors in each experimental
configuration invoked in this scheme. Moreover, a new scheme for
\emph{continuous} characterization of quantum dynamics can be
introduced, by utilizing weak measurements for the required quantum
error detections in DCQD \cite{Ahn02,Brun:05}. Finally, the general
techniques developed for direct characterization of quantum dynamics
could be further utilized for control of open quantum systems
\cite{mohseni-rezakhani-Aspuru07}.


\begin{acknowledgments}
We thank J. Emerson, D. F. V. James, K. Khodjasteh, A. T. Rezakhani,
A. Shabani, A. M. Steinberg, and M. Ziman for helpful discussions.
This work was supported by NSERC (to M.M.), and NSF Grant No. CCF-0523675, ARO Grant W911NF-05-1-0440, and the Sloan Foundation (to
D.A.L.).
\end{acknowledgments}

\begin{widetext}
\appendix
\section{Generalization to arbitrary open quantum systems}

\label{generalization-Algorithm}

Here, we first demonstrate that the overall measurements for a full
characterization of the dynamics of an $n$ qudit systems (with d
being a prime) become the tensor product of the required
measurements on individual qudits. One of the important examples of
such systems is a QIP unit with $r$ qubits, thus having a
$2^{r}$-dimensional Hilbert space. Let us consider a quantum system
consisting of $r$ qudits, $\rho =\rho _{1}\otimes \rho
_{2}\otimes \cdots \otimes \rho _{r},$ with a Hilbert space of dimension $%
N=d^{r}$. The output state of such a system after a dynamical map becomes $%
\mathcal{E} (\rho )=\sum_{m,n=0}^{N^{2}-1}\chi _{mn}~E_{m}\rho
E_{n}^{\dagger }$ where here $\{E_{m}\}$ are the unitary operator
basis elements of an $N$-dimensional Hilbert space. These unitary
operator basis elements can be written as
$E_{m}=X^{q_{m_{1}}}Z^{p_{m_{1}}}\otimes
X^{q_{m_{2}}}Z^{p_{m_{2}}}\otimes \cdots \otimes
X^{q_{m_{r}}}Z^{p_{m_{r}}}$ \cite{wootters04}. Therefore, we have:
\begin{eqnarray*}
\mathcal{E} (\rho ) &=&\sum_{m,n=0}^{N^{2}-1}\chi
_{mn}(X^{q_{m_{1}}}Z^{p_{m_{1}}}\otimes \ldots\otimes
X^{q_{m_{n}}}Z^{p_{m_{n}}})\rho _{1}\otimes \ldots\otimes \rho
_{n}(X^{q_{n_{1}}}Z^{p_{n_{1}}}\otimes \ldots\otimes
X^{q_{n_{r}}}Z^{p_{n_{r}}})^{\dagger } \\
&=&\sum_{m_{1},\ldots,m_{r},n_{1,},\ldots,n_{r}=0}^{d^{2}-1}\chi
_{(m_{1}\ldots m_{r})(n_{1}\ldots n_{r})}(E_{m_{1}}\rho
_{1}E_{n_{1}}^{\dagger })\otimes \ldots(E_{m_{s}}\rho
_{s}E_{n_{s}}^{\dagger })\ldots\otimes
(E_{m_{r}}\rho _{r}E_{n_{r}}^{\dagger }) \\
&=&\sum_{m_{1}\ldots m_{r},n_{1}\ldots n_{r}=0}^{d^{2}-1}\chi
_{(m_{1}\ldots m_{r})(n_{1}\ldots n_{r})}(E_{m}\rho E_{n}^{\dagger
})_{s}^{\otimes ^{r}},
\end{eqnarray*}%

\begingroup\squeezetable
\begin{table*}[tp]
\begin{ruledtabular}
\caption{Required physical resources for the QPT schemes: Standard
Quantum Process Tomography (SQPT), Ancilla-Assisted Process
Tomography using separable joint measurements (AAPT), using mutual
unbiased bases measurements (MUB), using generalized measurements
(POVM), see Ref.~\cite{mohseni-rezakhani-lidar07}, and Direct
Characterization of Quantum Dynamics (DCQD). The overall number of
measurements is reduced quadratically in the DCQD algorithm with
respect to the separable methods of QPT. This comes at the expense
of requiring entangled input states, and two-qudit measurements of
the output states. The non-separable AAPT schemes require many-body
interactions that are not available experimentally
\cite{mohseni-rezakhani-lidar07}.}
\begin{tabular}{lcccccccc}
 Scheme & $\text{dim}({\mathcal H})${\footnotemark[1]} & ${N}_{\text{inputs}}$ &
 ${N}_{\text{exp.}}${\footnotemark[3]} & measurements & required interactions \\
\colrule
SQPT & $d^n$ & $d^2n$ &  $d^{4n}$  & 1-body & single-body \\
AAPT & $d^{2n}$ & 1 &  $d^{4n}$  & joint 1-body & single-body \\
AAPT (MUB) & $d^{2n}$ & 1 & $d^{2n}+1$  & MUB & many-body \\
AAPT (POVM) & $d^{4n}$ & 1 &  $1$ & POVM & many-body \\
DCQD & $d^{2n}$ & $[(d+1)+1]^n$ & $d^{2n}$ & Stabilizer/Normalizer & single- and two- body \\
\end{tabular}
\label{tab-comp1}
\end{ruledtabular}
\footnotetext[1]{$\mathcal{H}$: the Hilbert space of each
experimental configuration} \footnotetext[3]{overall number of
experimental configurations}
\end{table*}
\endgroup

where we have introduced $E_{m_{s}}=X^{q_{m_{s}}}Z^{p_{m_{s}}}$ and
$\chi
_{mn}=\chi _{(m_{1},\ldots,m_{r})(n_{1},\ldots,n_{r})}$. I.e., $%
m=(m_{1},\ldots,m_{s},\ldots,m_{r})$ and
$n=(n_{1},\ldots,n_{s},\ldots,n_{r})$, and the index $s$ represents
a generic qudit. Let us first investigate the tensor product
structure of the DCQD algorithm for characterization of the diagonal
elements of the superoperator. We prepare the eigenstate of the
stabilizer operators $(E_{i}^{A}E_{j}^{B})_{s}^{\otimes ^{r}}$ and
$(E_{i^{\prime }}^{A}E_{j^{\prime }}^{B})_{s}^{\otimes ^{r}}.$ For
each qudit, the projection operators corresponding to outcomes
$\omega ^{k}$ and $\omega^{k^{\prime }}$ (where $k,k^{\prime }=0,1,\ldots,d-1),$ have the form $P_{k}=%
\frac{1}{d}\sum_{l=0}^{d-1}\omega ^{-lk}(E_{i}^{A}E_{j}^{B})^{l}$ and $%
P_{k^{\prime }}=\frac{1}{d}\sum_{l^{\prime }=0}^{d-1}\omega
^{-l^{\prime }k^{\prime }}(E_{i^{\prime }}^{A}E_{j^{\prime
}}^{B})^{l^{\prime }}.$ The
joint probability distribution of the commuting Hermitian operators $%
P_{k_{1}},P_{k_{1}^{\prime }},P_{k_{2}},P_{k_{2}^{\prime
}},\ldots,P_{k_{r}},P_{k_{r}^{\prime }}$ on the output state
$\mathcal{E} (\rho )$ is:
\begin{eqnarray*}
 \mathrm{Tr}[(P_{k}P_{k^{\prime }})_{s}^{\otimes ^{r}}\mathcal{E}
(\rho )] &=&\frac{1}{(d^{2})^{r}}%
\sum_{m_{1},\ldots,m_{r},n_{1},\ldots,n_{r}=0}^{d^{2}-1}\chi
_{(m_{1},\ldots,m_{r})(n_{1},\ldots,n_{r})}\times \\ &&
\left(\sum_{l=0}^{d-1}\sum_{l^{\prime }=0}^{d-1}\omega ^{-lk}\omega
^{-l^{\prime }k^{\prime }}\mathrm{Tr}[~E_{n}^{\dagger
}(E_{i}^{A})^{l}(E_{i^{\prime }}^{A})^{l^{\prime
}}E_{m}(E_{j}^{B})^{l}(E_{j^{\prime }}^{B})^{l^{\prime }}\rho
]\right)_{s}^{\otimes ^{r}}
\end{eqnarray*}
By introducing $E_{i}E_{m}=\omega ^{i_{m}}E_{m}E_{i}$ for each qudit
and using the relation $[(E_{i}^{A}E_{j}^{B})^{l}(E_{i^{\prime
}}^{A}E_{j^{\prime }}^{B})^{l^{\prime }}\rho ]_{s}=\rho _{s}$ we
obtain:
\begin{eqnarray*}
\mathrm{Tr}[(P_{k}P_{k^{\prime }})_{s}^{\otimes ^{r}}\mathcal{E}
(\rho )]
&=& \frac{1}{(d^{2})^{r}}\sum_{m_{1},\ldots,m_{r},n_{1},\ldots,n_{r}=0}^{d^{2}-1}%
\chi _{(m_{1},\ldots,m_{r})(n_{1},\ldots,n_{r})} \times
\left(\sum_{l=0}^{d-1}\sum_{l^{\prime }=0}^{d-1}\omega
^{(i_{m}-k)l}\omega ^{(i_{m}^{\prime }-k^{\prime })l^{\prime
}}\mathrm{Tr}[~E_{n}^{\dagger }E_{m}\rho ]\right)_{s}^{\otimes ^{r}}
\end{eqnarray*}
Using the QEC condition for nondegenerate codes, $%
\mathrm{Tr}[E_{n}^{\dagger }E_{m}\rho ]_{s}=(\delta _{mn})_{s},$ and
also using the discrete Fourier transform identities
$\sum_{l=0}^{d-1}\omega^{(i_{m}-k)l}=d\delta _{i_{m},k}$ and
$\sum_{l^{\prime }=0}^{d-1}\omega ^{(i_{m}^{\prime }-k^{\prime
})l^{\prime }}=d\delta _{i_{m}^{\prime },k^{\prime }}$ for each
qudit, we get:
\begin{eqnarray*}
\mathrm{Tr}[(P_{k}P_{k^{\prime }})_{s}^{\otimes ^{r}}\mathcal{E}
(\rho )]
&=&\sum_{m_{1},\ldots,m_{r},n_{1},\ldots,n_{r}=0}^{d^{2}-1}\chi
_{(m_{1},\ldots,m_{r})(n_{1},\ldots,n_{r})}(\delta _{i_{m},k}\delta
_{i_{m}^{\prime },k^{\prime }}\delta _{mn})_{s}^{\otimes ^{r}} \\
&=&\chi _{(m_{01},\ldots,m_{0r})(m_{01},\ldots,m_{0r})},
\end{eqnarray*}%
where for each qudit, the index $m_{0}$ is defined through the relations $%
i_{m_{0}}=k$ and $i_{m_{0}}^{\prime }=k^{\prime }$, etc. I.e.,
$E_{m_{0}}$ is the unique error operator that anticommutes with the
stabilizer operators of each qudit with a fixed pair of eigenvalues
$\omega ^{k}$ and $\omega ^{k^{\prime }}$corresponding to
experimental outcomes $k$ and $k^{\prime }$. Since $P_{k}$ and
$P_{k^{\prime }}$ operator have $d$ eigenvalues, we
have $d^{2}$ possible outcomes for each qudit, which overall yields $%
(d^{2})^{r}$ equations that can be used to characterize all the
diagonal
elements of the superoperator with a single ensemble measurement and $%
(2d)^{r}$ detectors. Note that in the above ensemble measurement we
can obtain $\log _{2}d^{4r}$ bits of classical information, which is
optimal according to the Holevo bound for an $2r$-qudit system of
dimension $d^{2}$. Similarly, the off-diagonal elements of
superoperators can be identified by a tensor product of the
operations in the DCQD algorithm for each individual qudit, see
Ref.~\cite{MasoudThesis}. A comparison of the required physical
resources for $n$ qudits is given in Table~\ref{tab-comp1}.

For a $d$-dimensional quantum system where $d$ is neither a prime
nor a power of a prime, we can always imagine another $d^{\prime
}$-dimensional quantum system such that $d^{\prime }$ is prime, and
embed the principal qudit as a subspace into that system. For
example, the energy levels of a six-level quantum system can be
always regarded as the first six energy levels of a virtual
seven-level quantum system, such that the matrix elements for
coupling to the seventh level are practically zero. Then, by
considering the algorithm for characterization of the virtual
seven-level system, we can perform only the measurements required to
characterize superoperator elements associated with the first six
energy levels.
\twocolumngrid
\end{widetext}



\end{document}